\documentclass[pdflatex,sn-mathphys-num]{sn-jnl}% Math and Physical Sciences Numbered Reference Style 
\usepackage{graphicx}% ,subcaption
\usepackage{multirow}%
\usepackage{amsmath,amssymb,amsfonts}%
\usepackage{amsthm}%
\usepackage{mathrsfs}%
\usepackage[title]{appendix}%
\usepackage{xcolor}%
\usepackage{textcomp}%
\usepackage{manyfoot}%
\usepackage{booktabs}%
\usepackage{algorithm}%
\usepackage{algorithmicx}%
\usepackage{algpseudocode}%
\usepackage{listings}%
\newcommand{\ruby}{RUBIES-EGS-QG-1\,}
\newcommand{\micron}{{\rm \mu m}}
\newcommand{\Nii}{[N\,{\sc ii}]}
\newcommand{\Ha}{H$\alpha$}
\newcommand{\Sii}{[S\,{\sc ii}]}
\newcommand{\Oiii}{[O\,{\sc iii}]}

\newcommand{\Msun}{\rm M_\odot}
\newcommand{\kms}{\rm km\,s^{-1}}
\newcommand{\Mpc}{\rm Mpc}
\newcommand{\Mhalo}{M_{\rm halo}}
\newcommand{\prospector}{\texttt{Prospector}\,}

\theoremstyle{thmstyleone}%
\theoremstyle{thmstyletwo}%

\theoremstyle{thmstylethree}%

\begin{document}

\title[Efficient formation of a massive quiescent galaxy at redshift 4.9]{Efficient formation of a massive quiescent galaxy at redshift 4.9}


\author*[1]{\fnm{Anna} \spfx{de} \sur{Graaff}}\email{degraaff@mpia.de}
\author[2,3]{\fnm{David J.} \sur{Setton}}
\author[4]{\fnm{Gabriel} \sur{Brammer}}
\author[5]{\fnm{Sam} \sur{Cutler}}
\author[6,7]{\fnm{Katherine A.} \sur{Suess}}
\author[8]{\fnm{Ivo} \sur{Labb\'e}}
\author[9]{\fnm{Joel} \sur{Leja}}
\author[10]{\fnm{Andrea} \sur{Weibel}}
\author[11]{\fnm{Michael V.} \sur{Maseda}}
\author[5,4]{\fnm{Katherine E.} \sur{Whitaker}}
\author[12]{\fnm{Rachel} \sur{Bezanson}}
\author[1]{\fnm{Leindert A.} \sur{Boogaard}}
\author[13,14]{\fnm{Nikko J. } \sur{Cleri}}
\author[15]{\fnm{Gabriella}\sur{De Lucia}}
\author[16]{\fnm{Marijn} \sur{Franx}}
\author[2]{\fnm{Jenny E.}\sur{Greene}}
\author[17,15]{\fnm{Michaela}\sur{Hirschmann}}
\author[18]{\fnm{Jorryt} \sur{Matthee}}
\author[19]{\fnm{Ian} \sur{McConachie}}
\author[20]{\fnm{Rohan P.} \sur{Naidu}}
\author[10,4]{\fnm{Pascal A.} \sur{Oesch}}
\author[12]{\fnm{Sedona H.} \sur{Price}}
\author[1]{\fnm{Hans-Walter} \sur{Rix}}
\author[21]{\fnm{Francesco} \sur{Valentino}}
\author[9]{\fnm{Bingjie} \sur{Wang}}
\author[22]{Christina C.\ Williams}

\affil*[1]{\small \orgname{Max Planck Institute for Astronomy}, \orgaddress{\street{K\"onigstuhl 17}, \city{Heidelberg}, \postcode{69117}, \country{Germany}}}

\affil[2]{\small \orgdiv{Department of Astrophysical Sciences}, \orgname{Princeton University}, \orgaddress{\street{4 Ivy Lane}, \city{Princeton}, \postcode{08544}, \state{New Jersey}, \country{United States}}}
\affil[3]{\small \orgname{Brinson Prize Fellow}}

\affil[4]{\small \orgdiv{Cosmic Dawn Center}, \orgname{University of Copenhagen}, \orgaddress{\street{Jagtvej 128}, \city{Copenhagen N} \postcode{2200}, \country{Denmark}}}

\affil[5]{\small \orgdiv{Department of Astronomy, University of Massachusetts}, \city{Amherst}, \state{MA} \postcode{01003}, \country{USA}}

\affil[6]{\small \orgname{Kavli Institute for Particle Astrophysics and Cosmology and Department of Physics, Stanford University}, \orgaddress{452 Lomita Mall}, \city{Stanford}, \postcode{94305}, \country{USA}}
\affil[7]{\small \orgname{NHFP Hubble Fellow}}

\affil[8]{\small Centre for Astrophysics and Supercomputing, Swinburne University of Technology, Melbourne, VIC 3122, Australia}

\affil[9]{\small \orgname{ Department of Astronomy \& Astrophysics, The Pennsylvania State University }, \city{University Park}, \postcode{16802}, \country{USA}}
%\affiliation{Institute for Computational & Data Sciences, The Pennsylvania State University}

\affil[10]{\small\orgdiv{Department of Astronomy}, \orgname{University of Geneva},
\orgaddress{\street{Chemin Pegasi 51}, \city{Versoix}, \postcode{1290},
 \state{GE}, \country{Switzerland}}}

\affil[11]{\small \orgdiv{Department of Astronomy}, \orgname{University of Wisconsin-Madison}, \orgaddress{\street{475 N. Charter St.}, \city{Madison}, \postcode{53706}, \state{Wisconsin}, \country{United States}}}

\affil[12]{\small \orgname{Department of Physics and Astronomy and PITT PACC, University of Pittsburgh}, \orgaddress{\city{Pittsburgh}, \state{PA}, \postcode{15260}, \country{USA}}}
 
\affil[13]{\small Department of Physics and Astronomy, Texas A\&M University, College Station, TX, 77843-4242 USA}

\affil[14]{\small George P.\ and Cynthia Woods Mitchell Institute for Fundamental Physics and Astronomy, Texas A\&M University, College Station, TX, 77843-4242 USA}

\affil[15]{\small\orgname{INAF - Astronomical Observatory of Trieste}, \orgaddress{\street{via G.B. Tiepolo 11}, \city{Trieste}, \postcode{34143}, \country{Italy}}}

\affil[16]{\small \orgname{Leiden Observatory, Leiden University}, \orgaddress{\street{P.O.Box 9513}, \postcode{NL-2300 AA} \city{Leiden}, \country{The Netherlands}}}

\affil[17]{\small\orgname{Institute for Physics, GalSpec laboratory, EPFL, Observatory of Sauverny}, \orgaddress{\street{Chemin Pegasi 51}, \city{Versoix}, \postcode{1290}, \country{Switzerland}}}

\affil[18]{\small \orgname{Institute of Science and Technology Austria (ISTA)}, \orgaddress{\street{Amp Campus 1}, \city{Klosterneuburg}, \postcode{3400}, \country{Austria}}}

\affil[19]{\small \orgname{Department of Physics \& Astronomy, University of California Riverside}, \city{Riverside}, \postcode{92521}, \country{USA}}

\affil[20]{\small \orgname{MIT Kavli Institute for Astrophysics and Space Research}, \orgaddress{\street{77 Massachusetts Ave.}, \city{Cambridge}, \postcode{02139}, \country{USA}}}

\affil[21]{\small \orgname{European Southern Observatory}, \orgaddress{\street{Karl-Schwarzschild-Str. 2}, \city{Garching bei M\"{u}nchen}, \postcode{85748}, \country{Germany}}}

\affil[22]{\small \orgname{NSF’s National Optical-Infrared Astronomy Research Laboratory}, \orgaddress{\street{950 North Cherry Avenue}, \city{Tucson}, \state{AZ} \postcode{85719}, \country{USA}}}


\abstract{Within the established framework of structure formation, galaxies start as systems of low stellar mass and gradually grow into far more massive galaxies\cite{White1978,Blumenthal1984}. The existence of massive galaxies in the first billion years of the Universe, suggested by recent observations, appears to challenge this model, as such galaxies would require highly efficient conversion of baryons into stars\cite{Labbe2023,BoylanKolchin2023,Behroozi2018,Xiao2023}. An even greater challenge in this epoch is the existence of massive galaxies that have already ceased forming stars\cite{Glazebrook2017,Valentino2020,Carnall2023b}. However, robust detections of early massive quiescent galaxies have been challenging due to the coarse wavelength sampling of photometric surveys. 
Here we report the spectroscopic confirmation with the James Webb Space Telescope of the quiescent galaxy \ruby\ at redshift $z=4.90$, 1.2 billion years after the Big Bang. Deep stellar absorption features in the spectrum reveal that the galaxy's stellar mass of $10^{11}\,\Msun$ 
formed in a short 200\,Myr burst of star formation, after which star formation activity dropped rapidly and persistently. According to current galaxy formation models, systems with such rapid stellar mass growth and early quenching are too rare to plausibly occur in the small area probed spectroscopically with JWST. Instead, the discovery of \ruby\ implies that early massive quiescent galaxies can be quenched earlier or exhaust gas available for star formation more efficiently than currently assumed.}

%\keywords{keyword1, Keyword2, Keyword3, Keyword4}



\maketitle

\begin{figure}
    \centering
    \includegraphics[width=0.95\linewidth]{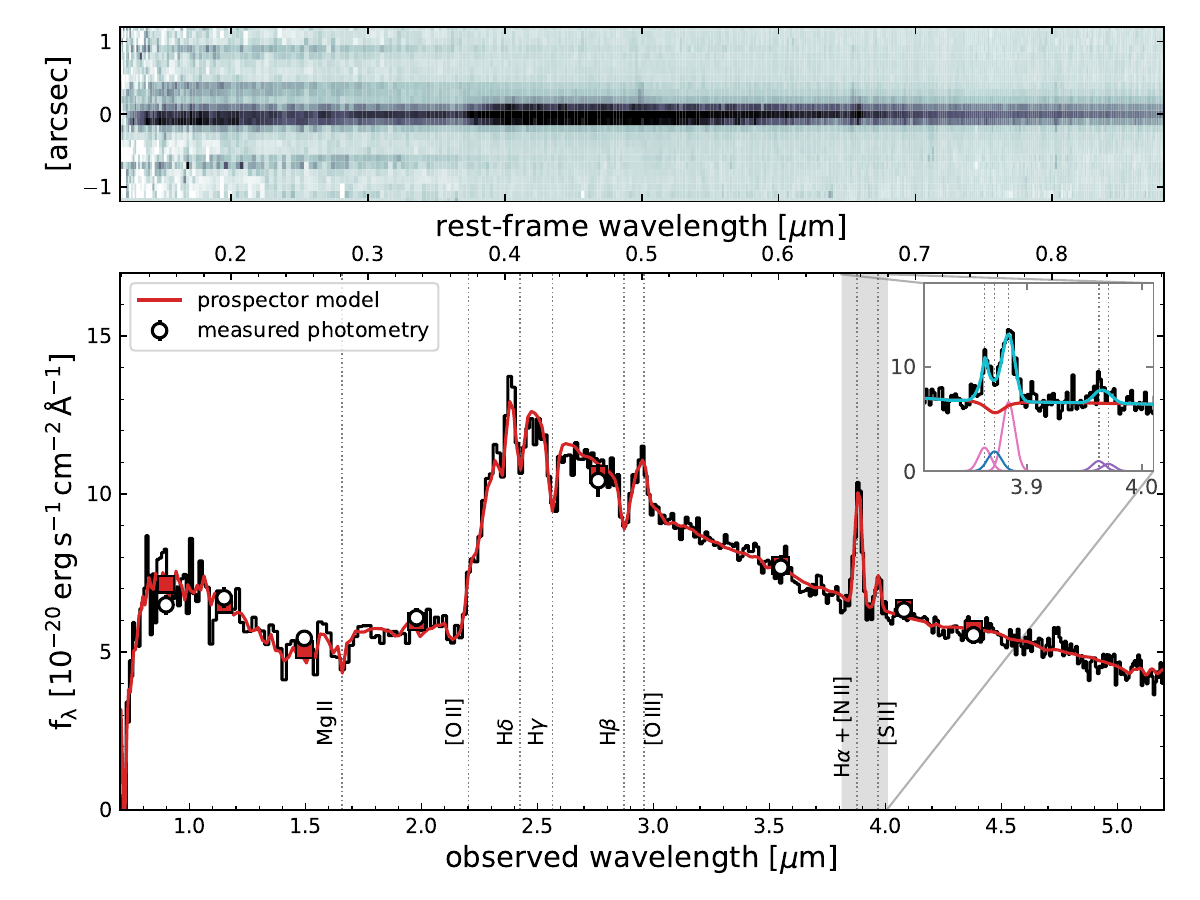}
    \caption{JWST/NIRSpec PRISM spectrum of the massive quiescent galaxy \ruby\ at a redshift of $z=4.8976$. The inset shows the medium-resolution (NIRSpec G395M) spectrum around the wavelength of \Ha. Both spectra were calibrated to the measured photometry using \prospector. The spectrum shows deep Balmer absorption lines, similar to post-starburst galaxies at lower redshifts, and implies a lack of star formation in its recent history. The presence of the emission line doublets \Oiii, \Nii, \Sii, and the minimal inferred infilling of the \Ha\ absorption line are consistent with AGN activity. 
    }
    \label{fig:prism_spec}
\end{figure}

\ruby\ was identified as a candidate massive quiescent galaxy at $z>4.5$ based on its red color measured from photometry across 1--5~$\micron$ in the Extended Groth Strip (EGS) obtained with the NIRCam instrument of the James Webb Space Telescope (JWST) \cite{Valentino2023,Carnall2023,Urbano2024}. The source was subsequently selected as a target for spectroscopic follow-up with JWST/NIRSpec %as part of the RUBIES program (JWST program GO-4233; PIs A. de Graaff and G. Brammer) 
because of its red color, $\rm F150W-F444W=2.35$, and bright apparent magnitude at long wavelengths, $\rm F444W=22.5$. The low-resolution PRISM spectrum of \ruby\ obtained with JWST/NIRSpec (Figure~\ref{fig:prism_spec}) reveals deep Balmer absorption lines and a strong spectral break at a rest-frame wavelength of $4000\,\AA$, indicating a lack of star formation in its recent history.

We detect the \Oiii$_{\lambda\lambda4960,5008}$, \Sii$_{\lambda\lambda6718,6733}$ and blended \Ha\ and \Nii$_{\lambda\lambda6549,6585}$ emission lines in the PRISM spectrum. A grating spectrum of \ruby\ with higher spectral resolution across 2.9--5.2~$\micron$
% and modeling of the emission lines (see Methods) 
reveals weak \Ha\ emission infilling the stellar absorption feature and strong \Nii\ emission {at a redshift of $z=4.8976\pm_{0.0010}^{0.0006}$} (see Methods). 
A marginal detection of \Ha\ implies {$\log {\rm [N\,{\sc II}]_{\lambda6585}/ {\rm H\alpha}} \approx 0.5$} and exceeds the ratio that can be explained by photoionization from massive stars by a factor {$\approx3$}\cite{Kewley2001}. Therefore, the emission lines do not appear to be connected to ongoing star formation activity, but rather suggest the presence of an active galactic nucleus (AGN), although we cannot rule out the presence of shocked gas\cite{Kewley2006}. Despite evidence for an AGN, the deep Balmer lines, indicative of a post-starburst system\cite{Goto2007}, suggest that the continuum emission of the spectrum is dominated by an evolved stellar population.

The high redshift of \ruby\ allows for stringent constraints on its star formation history in the first billion years of the Universe. To measure the star formation history, we use \prospector\cite{Johnson2021} to jointly fit a 21-parameter model to the PRISM spectrum and the observed photometry from the Hubble Space Telescope (HST) and JWST (see Methods for a complete description).
In brief, the model star formation history is parameterized as 14 time bins of constant star formation, with a common metallicity, dust attenuation, intrinsic velocity dispersion, and stellar initial mass function (IMF). We fit nebular emission lines using a simple model where lines are approximated by Gaussian profiles but no assumption is made on the origin of the emission, and utilize a {6th}-order polynomial to account for uncertainty in the flux calibration of the spectrum. We explore the effect of deviations in choices from our fiducial model in the Methods.

The median model of the sampled posterior is shown in red in Figure~\ref{fig:prism_spec}. We find a high stellar mass of {$9.9 \pm_{ 0.5 }^{ 0.4 }\times10^{10} \,\Msun$}, and a low star formation rate in the past 100 Myr,  {$\rm SFR_{100} = 4.0 \pm_{ 1.0 }^{ 3.5 }$\,$\Msun\,yr^{-1}$}. Together, these correspond to a low specific star formation rate of  {$\rm sSFR =  4.0\\pm_{0.8}^{1.0} \times 10^{-11}\,\rm yr^{-1}$} cementing \ruby as the highest-redshift spectroscopic confirmation of a massive quiescent galaxy to date. We infer low attenuation by dust, with the V-band attenuation from our modeling  {$A_{\rm v} =  0.17 \pm_{ 0.05 }^{ 0.06 }$}. 
The spectrum unambiguously demonstrates the red rest-frame optical color is dominated by old stars rather than dust-obscured star formation. This is corroborated by a non-detection in NOEMA observations at 1.1\,mm, which implies a $3\sigma$ upper limit on the dust-obscured star formation rate of $120\,\Msun\,{\rm yr}^{-1}$.

\begin{figure}
    \centering
    \includegraphics[width=0.95\linewidth]{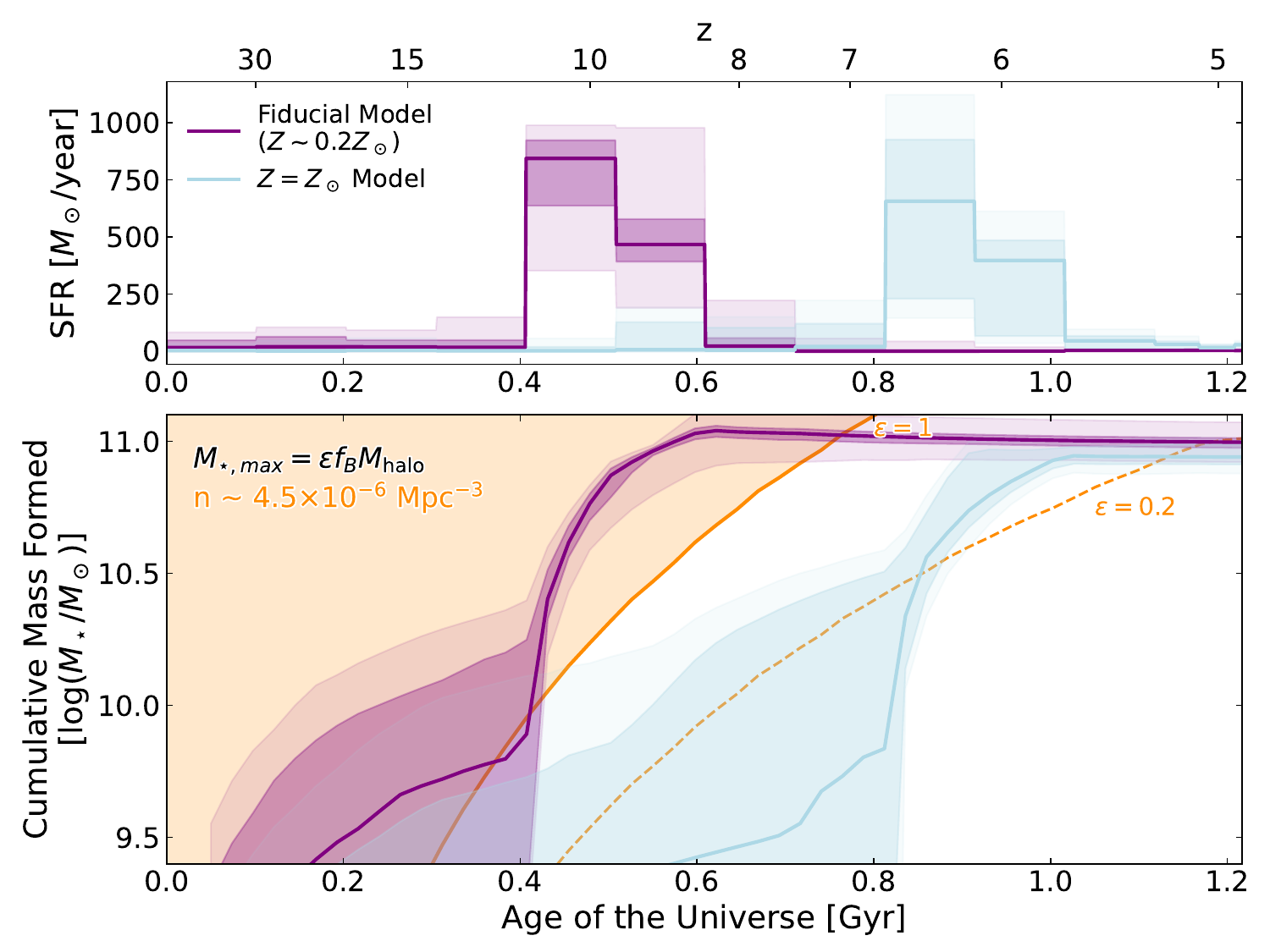}
    \caption{The history of stellar mass growth in \ruby. Top: Star formation history inferred from the modeling to the PRISM spectrum and photometry for the fiducial (free-metallicity) model (purple) and the fixed-solar metallicity model (blue). Dark (light) shaded regions indicate the $1\sigma$ ($2\sigma$) confidence intervals of the posterior distributions. 
    % Green lines indicate the redshift at which half the stellar mass was formed, $z_{form} =  9.0 \pm_{ 1.0 }^{ 1.1 }$ ($2\sigma$ uncertainties).
    Bottom: The cumulative mass history inferred from the star formation history of the two models. %, accounting for mass returned to the interstellar medium through stellar evolution.
    In orange we show the maximum stellar mass formed for a typical halo at the observed number density of massive quiescent galaxies at $z>4$\cite{Valentino2023}, assuming a universal baryon-to-total matter ratio ($f_{\rm B}$) and different baryon-to-stellar conversion factors ($\epsilon$). This indicates that a short ($\approx 300\,$Myr) burst of star formation with high efficiency of $\epsilon>0.2$ is required to form \ruby, corresponding to an efficiency at or greater than the peak of the stellar-halo mass relation\cite{Wechsler2018}. 
    }
    \label{fig:sfe}
\end{figure}

We show the resulting star formation history of the galaxy in the top panel of Figure~\ref{fig:sfe} in purple. \ruby\ assembled its stellar mass within a short burst of star formation, of {$\Delta t = 180\pm_{10}^{170}$\,Myr}, that peaked at a star formation rate of {$\rm SFR_{\rm peak}=870\pm_{140}^{70} \Msun\,yr^{-1}$}. For our fiducial model, half of the stellar mass was formed in the first  {$t_{\rm form} =  480 \pm_{ 10 }^{ 30 }$ Myr} of cosmic time, which would mark it as one of the earliest-forming massive galaxies observed. %We measure the time of quenching, $t_{90} =  746 \pm_{ 71 }^{ 120 }$ Myr. 
%Remarkably, 90\% of the galaxy's stellar mass, approximately the present-day stellar mass of Andromeda\cite{Tamm2012}, formed within just $t_{90} =  710 \pm_{ 31 }^{ 97 }$\,Myr after the Big Bang. 
We measure a corresponding quenching timescale  {$t_{90} - t_{\rm form} =  100 \pm_{ 10 }^{ 10 }$ Myr}, indicating that the decline in star formation from its peak in \ruby was extremely rapid.

Crucially, we find that the inferred stellar mass, the duration of the star formation burst, and the lack of recent star formation activity are {largely} robust against choices made in the model parameterization and star formation history priors. A full description of models tested can be found in the Methods. %Here we note that, even when assuming a rising SFH, which is expected in galaxies in the early Universe\cite{Papovich2011}, the posterior distributions still favor an early burst of star formation followed by rapid quenching, as the deep Balmer absorption features and lack of a UV continuum produced by young stars do not permit models with higher recent star formation activity. 
However, we find that the age of the stellar population depends strongly on the assumed metallicity of the system. {Our fiducial fit with the metallicity as free parameter indicates a low stellar metallicity ($\approx 0.2\,Z_\odot$), %consistent with the low metallicities found in massive quiescent galaxies at $z\sim1-2$\cite{Beverage2023}, 
and an old stellar population that formed as early as $z_{\rm form} =  10.0 \pm_{ 0.2 }^{ 0.4 }$ and stopped growing by $z\approx8.6$. Recent work has shown that the elemental abundance patterns in high-redshift quiescent galaxies differ from the solar abundance patterns typically used in stellar population modeling, which can lead to incorrectly inferred stellar metallicities \cite{Beverage2024}.} If we instead fix the metallicity of the stellar population to the solar value, we infer a substantially younger population with a formation redshift of $z_{\rm form}=6.3 \pm_{ 0.2 }^{ 0.1 }$ as well as more recent quenching (shown as the blue curve on Figure \ref{fig:sfe}). Although we cannot robustly differentiate between these two star formation histories with our current data, we stress that the stellar mass, the timescale and peak of the burst of star formation and quenching timescale do not depend significantly on metallicity. %However, this model is otherwise disfavored by both the PRISM data and the H$\alpha$ flux inferred from the G395M spectrum, and at odds with the low metallicities found in massive quiescent galaxies at $z\sim1-2$\cite{Beverage2023} and $z>4$\cite{Carnall2023} (see Section \ref{sec:sfh}). 

We estimate the dynamical mass of the system using the observed widths of the emission lines and the half-light radius measured from the F444W image, $r_{\rm e}=0.55\pm0.01\,$kpc (rest-frame wavelength of $0.75\,\micron$; see Methods): {we find that $\rm M_{dyn}= 2.7\pm_{0.8}^{0.7}\times10^{11}\,\Msun$ is consistent with \ruby\ being a very massive galaxy. The dynamical mass is a factor 3 higher than the stellar mass estimate, although we note that the estimated dynamical mass may be elevated by a factor $2-3$ due to non-gravitational motions of the ionized gas that we have not accounted for}. As the centers of high-redshift massive galaxies are expected to be dominated by stellar mass within the effective radius\cite{deGraaff2024}, this suggests that the stellar mass is not substantially under- or overestimated, despite uncertainties in the IMF assumed in our modeling. 

The stellar mass surface density within the estimated half-light radius of $\Sigma_*(<r_{\rm e})= 5.2^{+0.2}_{-0.3}\times10^{10}\,\Msun\,{\rm kpc}^{-2}$ is high, but well within theoretical limits for the maximum surface density achievable in a short burst of star formation\cite{Hopkins2010,Grudic2019}. %, whereas the more recently formed galaxy GS-9209 is close to this limit at $\Sigma_*(<r_{\rm e})\approx10^{11.15}\,\Msun\,{\rm kpc}^{-2}$. 
It is also in line with previous work at $z<3$ which demonstrated a strong link between low specific star formation rates and high central mass densities, with a threshold density for quiescence that increases toward higher redshift\cite{Franx2008,Whitaker2017}. This connection between galaxy structure and the star formation history has been interpreted as a compaction event followed by quenching due to feedback from intense star formation and an AGN\cite{Barro2013}, and is consistent with the high \Nii/\Ha\ ratio observed in \ruby\ that is indicative of an AGN or shocked gas.

With a peak star formation rate of $\approx 870\,\Msun\,\rm yr^{-1}$ at $z>6$, \ruby\ had a higher star formation activity than a large sample of the 40 most UV-luminous sources at $z\sim7$ discovered over an area of 7\,degree$^2$\cite{Bouwens2022}. Interestingly, the star formation rate and its star formation rate surface density ($\Sigma_{\rm SFR}\approx450\,\Msun\,\rm yr^{-1}\,kpc^{-2}$) are similar to the properties of the dusty star-forming galaxy G09 83808 at $z\approx6$ that resembles a local ultra-luminous infrared galaxy\cite{Zavala2018} { and the submillimeter galaxy SPT0311-58 at $z=6.9$ \cite{Strandet2017}. The star formation history of \ruby\ also matches well with the inferred star formation histories of the brightest red sources found with JWST \cite{Wang2024b} at $z\sim7$, provided that the light emitted by these sources originates from stars. These different observations suggest that the burst of star formation that formed \ruby\ could too have been strongly dust-obscured.} Such a link has also been suggested for the other spectroscopically-confirmed massive quiescent galaxy at $z>4.5$ \cite{Carnall2023b}, observed at $z=4.658$ with a stellar mass of $\approx3.8\times10^{10}\,\Msun$ formed at $z_{\rm form} =6.9\pm0.2$. In comparison with this source, however, \ruby\ stands out for being more than twice as massive. {On the other hand, a dust-obscured period of star formation may be difficult to reconcile with the inferred low metallicity {and the fact that to date no submillimeter galaxy has been found at $z>8$ with a star formation rate $>200\,\Msun\,{\rm yr^{-1}}$\cite{Bouwens2022,Schouws2022}. Therefore, this could instead} also imply that \ruby\ is an exceptionally rare source, or that the star formation history is either {significantly} more extended or more bursty than inferred from our modeling. In the latter scenario, finding {UV-luminous} progenitors at $z\sim 10$ may be difficult, as the probability of discovery depends on both the low number density of extremely massive high-redshift galaxies and on the duty cycle of star formation. }

The mere existence of \ruby\ provides an essential constraint on the growth of the most massive galaxies in the early Universe. In the bottom panel of Figure~\ref{fig:sfe} we show the cumulative mass assembly history of the galaxy derived from the modeled star formation history, subtracting mass returned to the interstellar medium through stellar evolution, as a function of the age of the Universe. %Based on the photometric selection of candidate massive quiescent galaxies in JWST imaging\cite{Valentino2023}, the comoving number density of quiescent galaxies of similar mass to \ruby\ ($M_*>10^{10.9}\,\Msun$) is low, approximately $4\times10^{-6}\,\Mpc^{-3}$ at $4<z<5$. 
{The small area targeted with JWST spectroscopy in the RUBIES survey ($\approx100$\,arcmin$^2$ thus far) implies an observed number density of $n\approx3\times10^{-6}\,\Mpc^{-3}$ at $4.5<z<5.5$. Based on the photometric selection of candidate massive quiescent galaxies in JWST imaging\cite{Valentino2023}, the comoving number density of quiescent galaxies of similar mass to \ruby\ ($M_*>10^{11}\,\Msun$) is approximately equally low, $n\approx4.5\times10^{-6}\,\Mpc^{-3}$ at $4<z<5$. }
%and based on the photometric selection of quiescent galaxies more massive than $M_*>10^{10.9}\,\Msun$ at $4<z<5$ we obtain a number density of $n\approx 5\times 10^{-6}\,\Mpc^{-3}$\cite{Valentino2023}.
We can hence use this number density to estimate the typical dark matter halo mass at a fixed redshift and derive an approximate limit on the total baryonic mass available within the halo\cite{BoylanKolchin2023}. This can then be converted to a maximum stellar mass at a given redshift by assuming $M_*(z)=\epsilon f_{\rm B} M_\mathrm{halo}(z)$, where $f_{\rm B}$ is the cosmic baryon fraction (15.6\%, \cite{Planck2020}) and $\epsilon$ is the baryon-to-star conversion efficiency. We plot this expected stellar mass for two values of the baryon efficiency: $\epsilon=1$ (i.e. assuming total conversion of baryons to stars), 
and the efficiency at the peak of the stellar-halo mass relation of $\epsilon=0.2$\cite{Wechsler2018}. The cumulative mass assembly history of \ruby\ implies a high efficiency of star formation of $\epsilon>0.2$, and the low-metallicity model even suggests that the galaxy is converting baryons to stars with near-perfect efficiency. %closely resembles the $\epsilon=1$ curve: if the extremely rapid mass assembly occurred in situ, this indicates that the galaxy is either converting baryons to stars with near-perfect efficiency, or resides in an unusually massive halo. 

Neither the extremely rapid mass assembly nor the early quenching of \ruby\ are consistent with predictions from large-volume 
($200^3-800^3\,\Mpc^3$) cosmological hydrodynamical (FLARES\cite{Lovell2023}, Magneticum Pathfinder\cite{Kimmig2023} and TNG300\cite{Hartley2023}) and semi-analytic (GAEA\cite{DeLucia2024}, SHARK\cite{Lagos2024}) simulations of galaxy formation. While some of these models are able to produce galaxies that are quiescent at $z\sim4.5-5.0$ and as massive as \ruby, such systems are extremely rare: the comoving number densities are approximately $1-10\times10^{-8}\,\Mpc^{-3}$ in the different simulations, and correspond to very massive haloes. In comparison with the estimated number density of \ruby, this implies a probability of 1\% ($2\sigma$ outlier) that such a source is observed in the small area probed spectroscopically with JWST ($\sim 100\,$arcmin$^2$ for the RUBIES program). However, at higher redshifts the comoving number density of quiescent galaxies with stellar massses $M_*>10^{11}\,\Msun$ decreases dramatically, with most simulations containing zero such galaxies at $z\geq5$. The inferred star formation history of \ruby, where star formation is quenched at $z\gtrsim5.5$, therefore implies that it would be a significant outlier at earlier times.

This indicates that the star formation and feedback recipes in the simulations do not accurately capture the formation and quenching processes of early massive galaxies. Alternatively, to reconcile the formation history of \ruby\ with simulations requires that our estimated observed number density is substantially overestimated, and that the galaxy instead resides in a very rare, and therefore massive, halo ($\Mhalo\gtrsim10^{13}\,\Msun$ by $z\sim5$). The effects of cosmic variance may be large for the relatively small area targeted spectroscopically with JWST: although unlikely, it is possible that the observation of \ruby\ is a $\gtrsim 3\sigma$ chance finding and indeed resides in a very massive halo.

\begin{figure}
    \centering
    \includegraphics[width=0.95\linewidth]{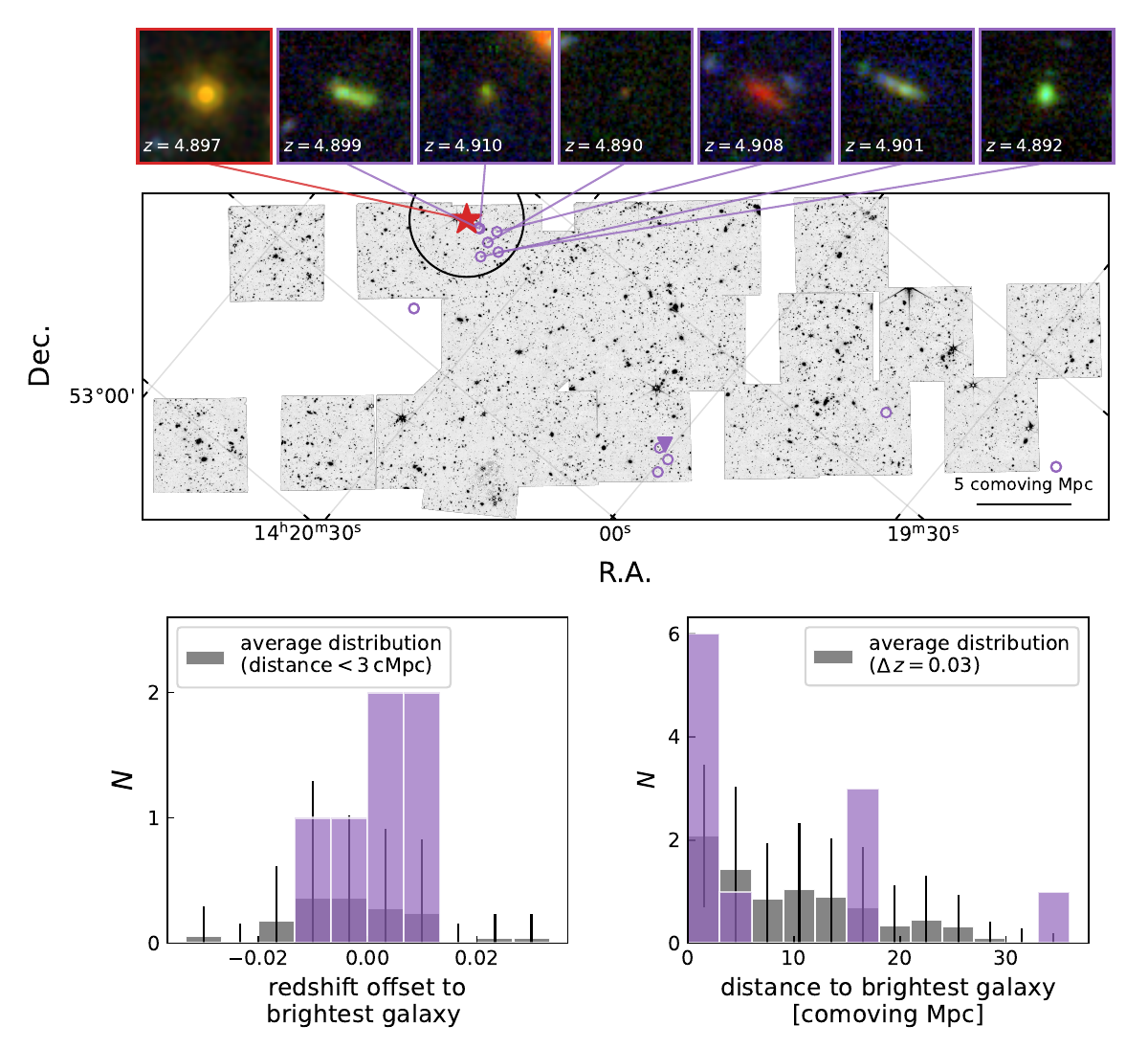}
    \caption{Spatial clustering of spectroscopically-confirmed sources at $z\approx4.90$ (circles) around \ruby\ (red star). The submillimeter galaxy, the brightest source among the group of 4 at a projected distance of 16 comoving Mpc, is indicated by a purple triangle. The background image shows the NIRCam F444W mosaic of the EGS field. Compared to the typical projected spatial clustering (within redshift ranges $\Delta z=0.03$) and redshift clustering (within apertures of radius $<3\,\Mpc$) for galaxies in the redshift range $4<z<6$ with robust redshifts from JWST spectroscopy, \ruby\ clearly resides in an overdense environment, forming the highest redshift known overdensity hosting a massive quiescent galaxy. We show false-color images (created from NIRCam F150W, F277W and F444W images) of \ruby\ and its 6 nearby neighbors. }
    \label{fig:environment}
\end{figure}

We examine the environment of \ruby\ for evidence of such a large-scale overdensity by compiling all sources in the EGS with robust redshifts from JWST spectroscopy, obtained from a mixture of JWST Cycle 1 and 2 programs using the DAWN JWST Archive (see Methods). In total, we find 6 sources in the direct vicinity of \ruby\ with {a redshift separation $\Delta z<0.013$} and projected separation of $<1$ arcmin ($<2$ comoving Mpc), and another 7 sources spread across the field at the same redshift but with larger angular separation (Figure~\ref{fig:environment}). Notably, we find a clustering of 4 sources at a distance of $\approx 16\,$comoving Mpc from \ruby, of which the brightest galaxy (measured at $4\,\micron$) is a sub-millimeter galaxy identified previously in SCUBA-2 data\cite{Zavala2017}. In comparison with the average density of the $4<z<6$ galaxy population, we find clear evidence for an overdensity at $z=4.90$ within an aperture of $\pi\times 3\,\Mpc^2$ (as also suggested in previous literature\cite{Arrabal2023,Naidu2022}), forming the highest redshift overdensity containing a massive quiescent galaxy found thus far\cite{Kakimoto2024}. However, based on these data alone, we cannot conclusively determine whether the region indeed represents an extremely massive halo, and whether the other 7 sources are associated with the overdensity.

The fact that the galaxy resides in an overdense environment may also point to substantial ex-situ mass accretion, in addition to in-situ star formation. 
A major merger between two approximately equally massive systems would provide a rapid accretion of stellar mass, allowing for more conventional star formation efficiencies. Moreover, major mergers have been proposed as a quenching mechanism\cite{DiMatteo2005}.  
However, equal-mass mergers between massive galaxies are exceptionally rare at $z>5$, as the simulations predict a number density of $\lesssim10^{-8}\,\Mpc^{-3}$ for such systems. 
We also do not find signatures of recent merging in the morphology of \ruby\ (Figure~\ref{fig:environment}; Methods).

Clearly, the rapid assembly of \ruby\ and its early quenching requires an extreme formation scenario. In the context of recent studies that have reported candidate massive galaxies at $z>6$\cite{Labbe2023} and massive quiescent galaxies with formation times at $z\gtrsim6$\cite{Nanayakkara2024,Carnall2023,Glazebrook2023,AntwiDanso2023}, \ruby\ stands out for its very high stellar mass, high redshift and deep Balmer absorption features that unambiguously set its formation at $z>6$. 
Although theoretical models can form extremely massive galaxies at early epochs, and some also produce quiescent galaxies at $z>5$, the number densities of these sources are extremely low. In the case of \ruby, this may indicate that the galaxy lies in an extremely rare, massive halo for its redshift, or the exceptional detection of a major merger between two massive galaxies. However, both of these scenarios are expected to be very rare ($\lesssim 0.1$ per degree$^2$), and the area covered by JWST spectroscopy thus far is small. The direct implication of the existence of \ruby\ is therefore that the star formation and feedback prescriptions in theoretical models require revision, as the model universes currently cannot reproduce the stellar mass growth and early quenching required to match the inferred abundance of massive quiescent galaxies.

\clearpage
\section*{Methods}\label{sec11}

\section{Spectroscopic data}\label{sec:nirspec}

The RUBIES program (GO-4233; PIs A. de Graaff and G. Brammer) is a JWST Cycle 2 program using the NIRSpec microshutter array (MSA)\citep{Ferruit2022} to observe galaxies in the CANDELS EGS and UDS extragalactic deep fields \citep{Candels1,Candels2}. Specifically, RUBIES targets galaxies detected in F444W from JWST/NIRCam imaging in the Cosmic Evolution Early Release Science (CEERS; program \#1345; PI S. Finkelstein) and Public Release IMaging for Extragalactic Research (program \#1837; PI J. Dunlop) surveys, and is optimized to reach high spectroscopic completeness for bright and red sources at $z>3$. Thus far, the survey has targeted an area of approximately $\rm 100\,arcmin^2$. Details about the target selection and prioritization is described in \cite{deGraaff2024d}.

\ruby\ (R.A. 214.9155459, Dec. 52.9490183) was observed in March 2024 as part of the observations in the EGS field. The MSA pointings were observed for 48 minutes each in the PRISM/CLEAR and the G395M/F290LP spectroscopic modes. Each target was observed in a 1$\times$3 configuration of open microshutters, with a 3-point nodding pattern. The spectra were allowed to overlap on the detector in the G395M exposures as most sources have faint enough continua to not severely contaminate other spectra.

The NIRSpec data are reduced using the \texttt{msaexp}\cite{Brammer2022} pipeline version 3{, the details of which are described in \cite{deGraaff2024d}. Briefly, in comparison to version 2 of the pipeline described in \cite{Heintz2024} we use updated reference files for improved flux calibration. In addition, we leverage empty sky shutters from the RUBIES program to derive custom bar shadow corrections, which we find provide a substantial improvement over the default reference files, removing strong ($\sim 10\%$ level) unphysical discontinuities in the extracted spectrum. We also use the empty sky shutters to construct a global background subtraction for the PRISM spectra. We use local background subtraction from the nodded exposures for the G395M spectrum, as the overlapping traces on the detector in this mode make it difficult to construct a global background solution. A Gaussian profile is fit to the 2D PRISM spectra to estimate the intrinsic width and centroid of trace. The 1D spectra for both dispersers are then extracted using this Gaussian profile with an optimal weighting\cite{Horne1986}.} We scale-up the 1-$\sigma$ errors on the 1D extracted spectrum by a factor of 1.7 to account for under-estimated uncertainties when comparing the pixel-to-pixel variations with the \texttt{msaexp}-derived errors \cite[as described in][]{Maseda2023}. The final flux calibration of the \ruby\ spectrum is performed by matching the continuum level in the PRISM to the multi-band photometry from HST and JWST/NIRCam, as described in the next Section.

\section{Photometric data}\label{sec:phot}

% \begin{itemize}
%     \item Photometry from v7.2 of Gabe's grizli reductions (ref to Valentino+ 2023 for details)
%     \item Filters: F435W, F606W, F814W (HST ACS); F125W, F140W, F160W (HST WFC3); F115W, F150W, F200W, F277W, F356W, F444W (JWST NIRCam)
%     \item mention custom background subtraction for F814W image to deal with the diffraction spike
%     \item photometry from PSF-matched images -> scaled to total flux based on Sersic fit in F444W
% \end{itemize}

{We use publicly available JWST/NIRCam imaging from the CEERS program\cite{Bagley2023} as well as program GO-2234 (PI: Ba\~nados; Khusanova et al. in prep.), which combined provide 8 bands of photometry (F090W, F115W, F150W, F200W, F277W, F356W, F444W and F410M).} %{In addition, we use archival HST/WFC3 imaging in 3 bands of photometry (F125W, F140W, and F160W) obtained as part of the CANDELS survey\cite{Candels1,Candels2}.}

We use the reduced image mosaics from the DAWN JWST Archive {(DJA; version 7.4)}. All images were reduced using \texttt{grizli} \citep{grizli} and have a pixel scale of $0.04\,$arcsec$\,{\rm pix}^{-1}$ (see also \cite{Valentino2023} for further details on the reduction). Next, we use empirical point spread function models (PSF) to construct mosaics that are PSF-matched to the F444W mosaic, as described in \cite{Weibel2024}. We measure fluxes in circular apertures with a radius of $0.25$\,arcsec from the PSF-matched photometry, centered on the centroid position estimated by running SourceExtractor\cite{Bertin1996} on an inverse-variance weighted stack of the F277W, F356W and F444W bands. 
%We note that \ruby\ is located in the vicinity of a bright star, which strongly affects the photometry in the HST/F814W image. For this band, we construct a custom background model to subtract the diffraction spike from the image prior to measuring aperture photometry. Finally, we rescale all aperture fluxes by the ratio of the total flux estimated from the best-fit S\'ersic profile in the F444W band (described in Section~\ref{sec:size}) and the F444W aperture flux.

\begin{figure}
    \centering
    \includegraphics[width=0.7\linewidth]{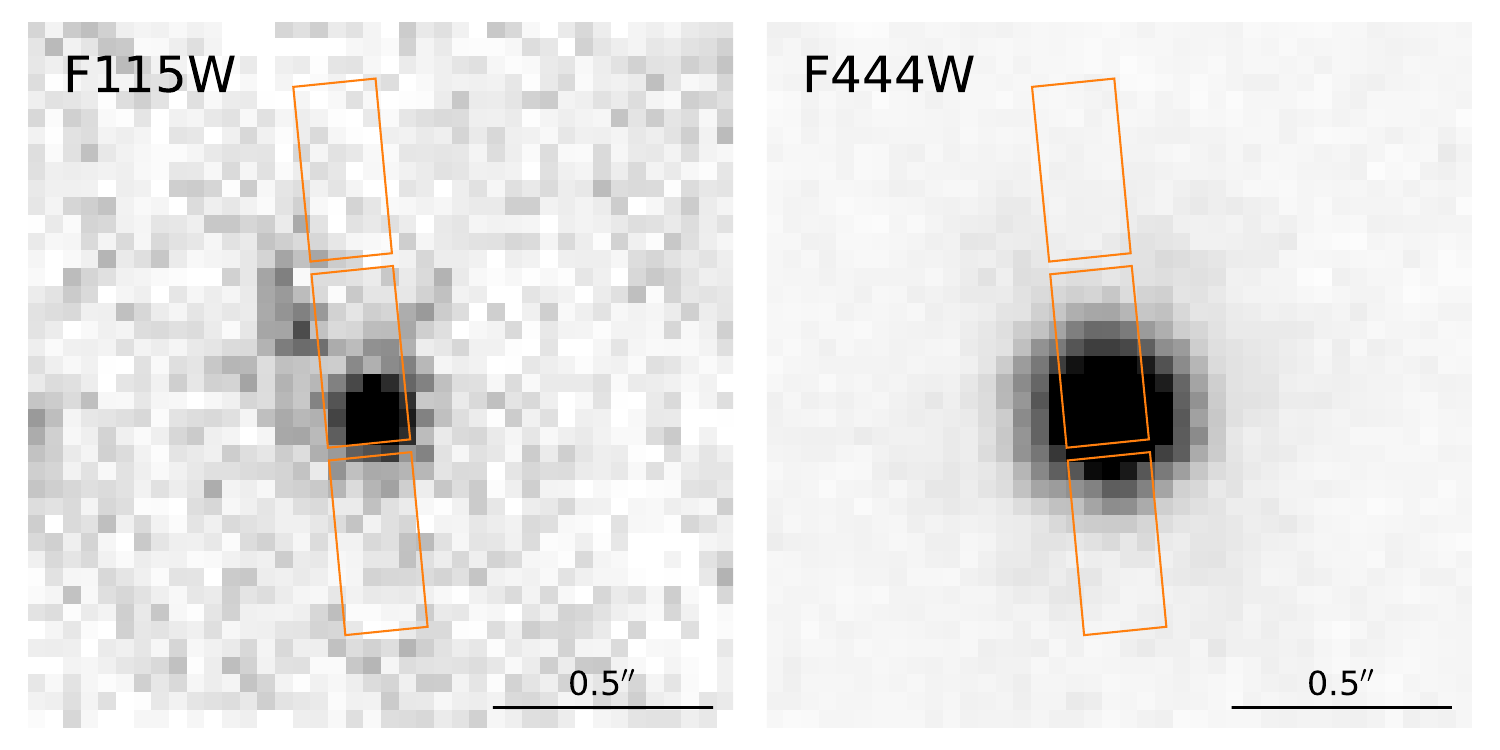}
    \caption{{NIRCam F115W and F444W image cutouts of \ruby. Orange lines show the location of the NIRSpec microshutters.}}
    \label{fig:image_stamps}
\end{figure}

{We show image cutouts of the F115W and F444W NIRCam filters in Figure~\ref{fig:image_stamps}, and include the position of the NIRSpec microshutters. The centroid of \ruby\ is located in the bottom of the microshutter, with some of the light falling on the bar between two microshutters. Due to the large variation in the PSF width as a function of wavelength, the slit losses introduced by the bar shadow are highly complex, the effect of which we discuss further in Section~\ref{sec:prospector}. We also find a faint blue clump in the F115W image, located $\approx0.2$ arcsec from \ruby. In Section~\ref{sec:eline_fitting} we show that emission lines from this source reveal that it is a satellite with a velocity separation of $\approx 600\,\kms$.}

\section{SED modeling}\label{sec:prospector}

\begin{figure}
    \centering
    \includegraphics[width=0.95\textwidth]{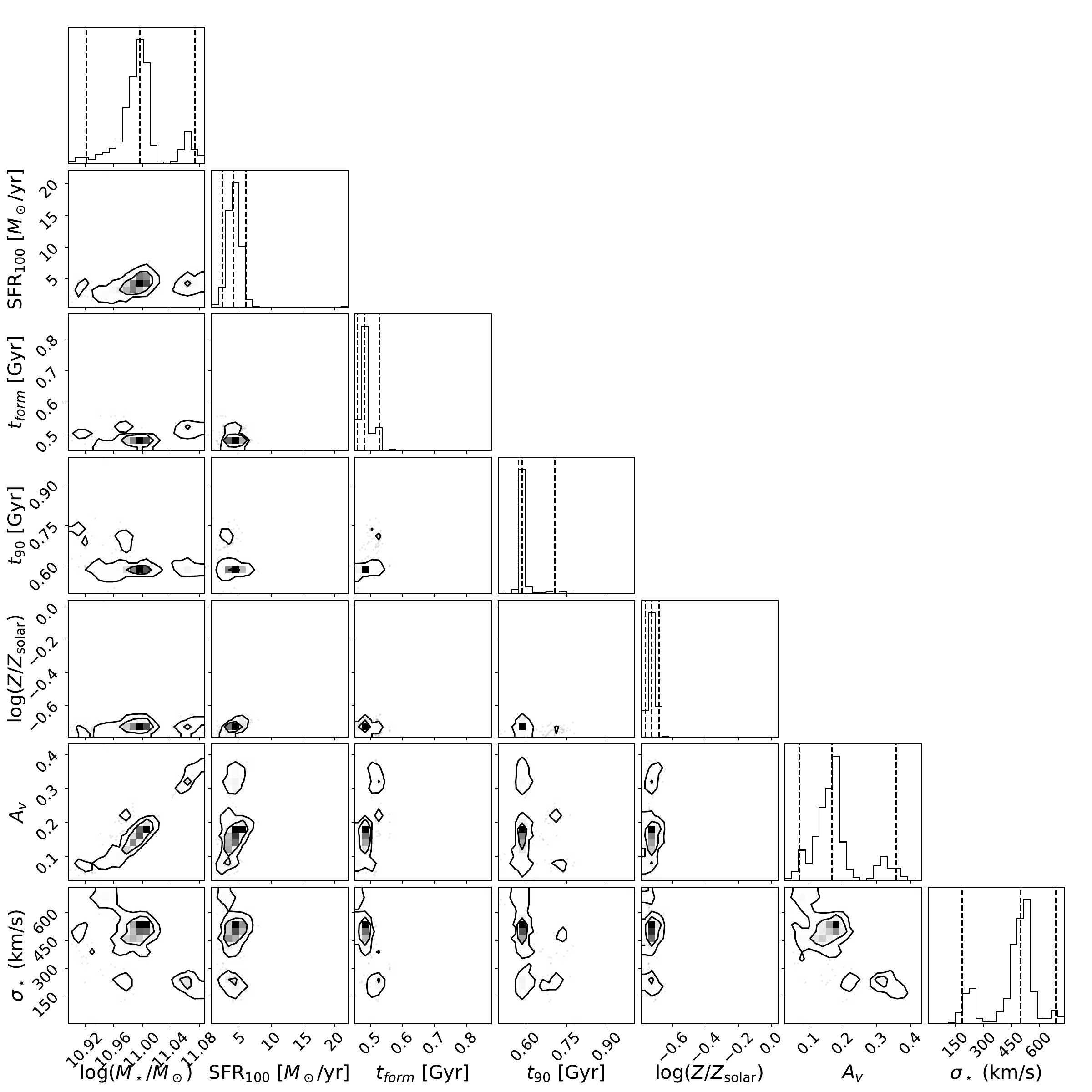}
    \caption{The covariant posteriors for a selected set of parameters in our fiducial fit with \prospector, with contours bounding 68\% and 95\% of the likelihood and dashed lines capturing the 95\% confidence interval on the marginalized posteriors. }
    \label{fig:corner}
\end{figure}

In order to measure the stellar population properties of \ruby, we utilize the Bayesian SED fitting code \prospector \cite{Johnson2017,Leja2017,Johnson2021} to fit non-parametric star formation histories to the NIRSpec/PRISM spectrum and the JWST/NIRCam photometry of this galaxy. We enforce a signal-to-noise ceiling of 20 on our photometric measurements to account for systematic uncertainties in the underlying stellar population models. We utilize the Flexible Stellar Population Synthesis (FSPS) stellar population synthesis models \cite{Conroy2009, Conroy2010}, the MILES spectral library \cite{Sanchez-Blazquez2006}, and MIST isochrones \cite{Choi2016,Dotter2016}. We assume a Chabrier initial mass function \cite{Chabrier2003} and fix the model redshift to $z_{\rm prism}=4.906$, the PRISM spectroscopic redshift estimated with \texttt{msaexp}, which differs slightly from the redshift derived from the G395M spectrum due to wavelength calibration uncertainties between the NIRSpec dispersers\cite{Bunker2023}. {We mask the outer edges of the spectrum, to avoid uncertainties in modeling absorption from the intergalactic medium (i.e. blueward of rest-frame $1200\,\AA$) and the absolute flux calibration at the edge of the NIRSpec CLEAR filter where we do not have photometric coverage ($>5.1\,\mu$m).}

Our fiducial star formation history parameterization is a 14-bin non-parametric model utilizing the \prospector continuity prior, with the logarithmic ratio between neighboring bins fit with a Student's t-distribution prior centered 0 with a width of 0.3 and $\nu=2$ following \cite{Leja2019}. We divide the {most recent} 100 Myr of star formation into three bins of width 5, 25, and 75 Myr respectively to provide fine sampling of the most recent star formation history, and fill the remaining age of the universe with {eleven} linearly spaced 100 Myr bins. We assume a two-parameter dust law with free $A_v$ and dust index spanning [0,2.5] and [-1,0.4] respectively \cite{Kriek2013}, and we fix the attenuation around young ($t<10^7$ Myr) stars to be twice that of the older populations. We fit for a free logarithmically sampled stellar metallicity in the range of [0.1$Z_\odot$, 2$Z_\odot$]. We perform sampling using the \texttt{dynesty} nested sampling package \cite{Speagle2020}.

\begin{figure}[h!]
    \centering
    \includegraphics[width=0.95\textwidth]{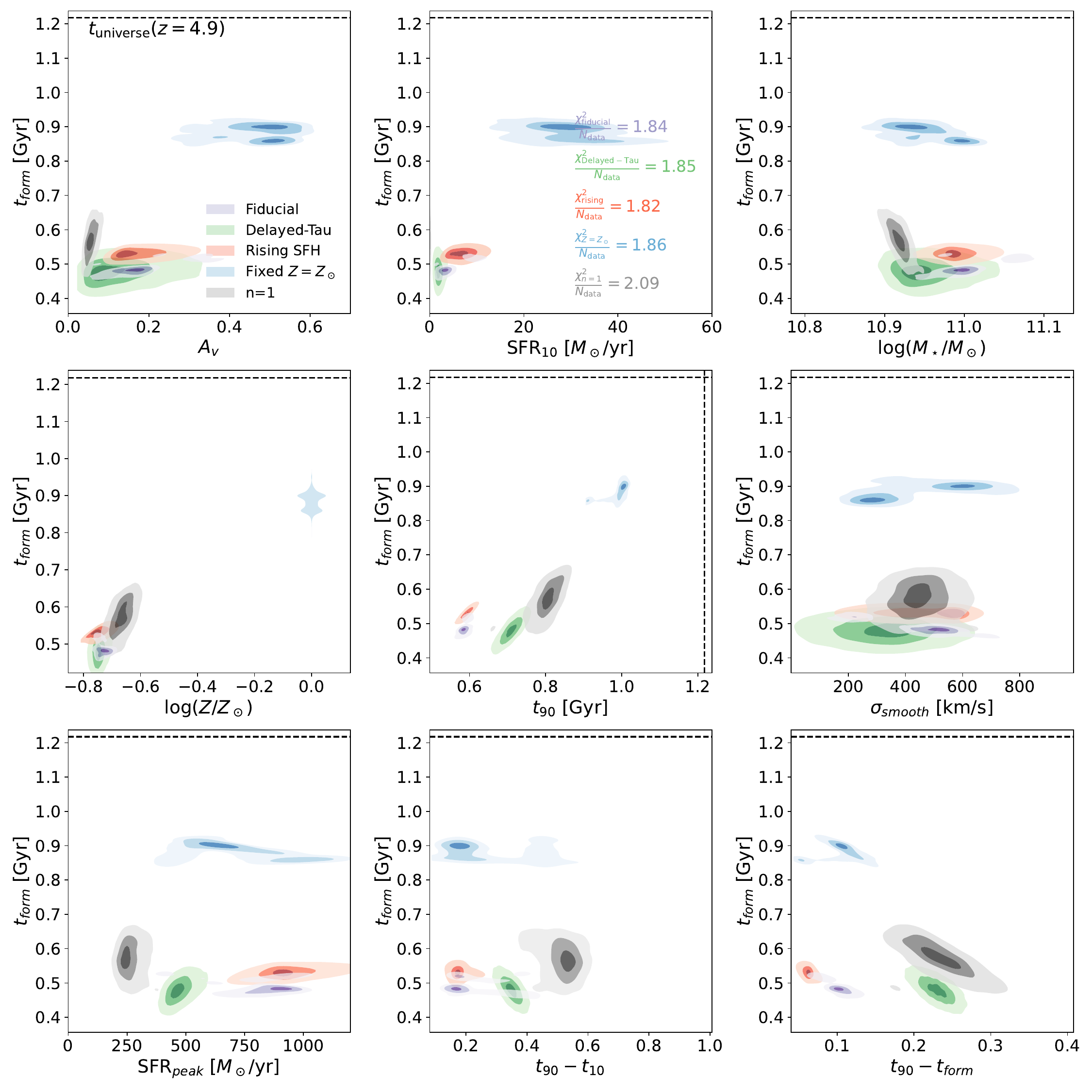}
    \caption{A comparison of the posteriors of our fiducial (purple), delayed-tau (green), our rising SFH prior (red), fixed $Z=Z_\odot$ (blue), and n=1 polynomial (grey) models, showing the age of the universe when 50\% of the galaxy's mass formed, $t_{\rm form}$, against the covariant parameters of $A_V$ (top left), star formation rate (top center), stellar mass (top right), metallicity (middle left), $t_{90}$, (the age of the universe when 90\% of the galaxy's mass formed, middle center), $\sigma_{\rm smooth}$ (middle right), SFR$_{\rm peak}$ (bottom left), star formation duration timescale ($t_{90}-t_{10}$, bottom middle), and star formation decline timescale ($t_{90}-t_{\rm form}$, bottom right). Black dashed lines indicate the age of the universe at the time of observation. The fiducial fit, the rising SFH prior fit, and the delayed-tau fits generally agree in reaching old, low-SFR, low-dust, metal-poor solutions. However, fits where we require that the average stellar metallicity to be solar result in a later-forming, more star forming, and dustier galaxy. In the bottom row, we highlight that even when metallicity is fixed to solar, the constraints on the peak star formation rate and the need for a rapid period of intense star formation are consistent with the earlier-forming fiducial model.}
    \label{fig:model_comp}
\end{figure}

In order to account for the NIRSpec/PRISM resolution, we convolve all models using the JDOX PRISM resolution curve scaled by a multiplicative factor of 1.3 as in \cite{CurtisLake2023}, which approximates the line spread function of a compact source. As \ruby is quite massive, we expect there to be significant intrinsic broadening of the stellar continuum due to the random motions of stars. As such, we fit for an additional free continuum velocity dispersion, with a deliberately large prior ($\sigma_\mathrm{smooth}=[0,1000]\,\kms$) that can also account for uncertainty in the precise normalization of the NIRSpec line spread function. We additionally account for uncertainty in the NIRSpec flux calibration by fitting using the \prospector\ \texttt{PolySpecModel} prescription, which marginalizes out a {6th} order {multiplicative} polynomial in order to rectify the observed spectrum to the model during each likelihood call. The choice to calibrate out such a high-order polynomial was motivated by the significant wavelength baseline ($\sim5 \ \mu$m) of the PRISM {spectrum} coupled with the considerable uncertainty in the flux calibration on small scales due to {the effect of differential slit losses. Because the source is located partially on the bar between two shutters (Figure~\ref{fig:image_stamps}) and the PSF of JWST depends strongly on wavelength, the effect of the bar shadows also has a strong wavelength dependence. Our empirical bar shadow correction (Section~\ref{sec:nirspec}) derived from blank sky shutters provides only a first order correction of this effect, as the sky emission fills the slit uniformly, but the morphology of \ruby\ follows a steep S\'ersic profile (Section~\ref{sec:size}). } 
This conservative approach utilizes the spectrum only for sharp spectral features such as emission or absorption lines and spectral breaks, and relies on the better-calibrated photometry to fix the shape of the spectral energy distribution. We note that using a lower-order polynomial (e.g. $n=1$) {yields a similarly old, low-metallicity stellar population (also shown in Figure~\ref{fig:model_comp}), although with a marginally lower stellar mass and more extended star formation history than our fiducial fit. However, we find that this fit has a significantly higher $\chi^2$ value ($\chi^2_{n=1}-\chi^2_{n=6}\approx 100$) and shows oscillatory features in the residuals, indicative of flux calibration issues.}

Finally, we account for nebular emission in key emission lines (Ly$\alpha_{1216}$, [OII]$_{\lambda\lambda3727,3729}$, [OIII]$_{\lambda\lambda4960,5008}$, H$\delta_{\lambda4103}$, H$\gamma_{\lambda4342}$, H$\beta_{\lambda4864}$, H$\alpha_{\lambda6564}$, [NII]$_{\lambda\lambda6549,6585}$, and [SII]$_{\lambda\lambda6718,6733}$) using the \prospector nebular marginalization procedure, fitting for emission lines in the residual spectrum using a least squares algorithm and incorporating those models in each likelihood call. This procedure allows the emission lines to be produced without ascribing the emission to a specific source; this is important given the strong evidence for non-star formation sources of ionizing radiation (see Section~\ref{sec:eline_fitting}), although our conclusions are unchanged if we do allow for physical nebular infilling of the Balmer absorption lines. {Similar to the stellar continuum, we convolve all emission lines with a free emission line velocity dispersion $\sigma_\mathrm{gas}=[0,1000]\,\kms$, where we assume all emission lines have the same width.} In Figure \ref{fig:corner}, we show the covariant posteriors for key parameters characterizing \ruby and demonstrating that the fits are well converged.

\section{Star formation history testing}\label{sec:sfh}

The central result of the star formation history fitting of \ruby is the finding that the galaxy assembled its high ($\sim10^{11}\,\Msun$) stellar mass and ceased forming stars within only the first billion years of cosmic time, straining galaxy evolution models. 
Here, we explore how different choices about the parameterization of the star formation history or our priors could impact the inferred age of this system.

First, we examine the impact of choosing a different parameterization of the star formation history. Non-parametric models for the star formation history are commonly chosen because of their flexibility to account for a wide range of star formation history shapes, but such parameterizations systematically measure older ages with increased uncertainty \cite{Leja2019_3DHST}. As a soundness check, we perform our fiducial fit with a star formation history parameterized with the commonly utilized delayed-$\tau$ ($\mathrm{SFR} \propto \ t e^{-t/\tau}$) with just three free parameters describing the star formation history, rather than fourteen. We find that this parametric star formation history almost identically recovers that of the non-parametric model, measuring $t_{\rm form} = 480\pm_{40}^{30}$ and recovering similar posteriors for the star formation rate, stellar mass, metallicity, and $t_{90}$ (see Figure \ref{fig:model_comp} and Table \ref{tab:sfh_results}), {albeit with a slightly longer $t_{90} - t_{10}$}.

Second, we test our non-parametric star formation history against the choice of prior to mitigate the concern that the inferred star formation history is prior-driven \cite{Leja2019,Wang2024:sys}. Instead of the flat prior assumed in our fiducial setup, we impose a rising star formation history prior, as expected from e.g., \cite{Papovich2011,Behroozi2019}, by shifting the mean logarithmic ratio of our Student's-t prior from 0.0 to 0.3 and re-running our fiducial fit. We find that our measurement of $t_{\rm form}$ is largely insensitive to the imposition of this prior, and the galaxy is still fit as $\sim550$ Myr old, in spite of a slight increase in implied star formation rate and formation time. 

We find that both the fiducial fit (insensitive of SFH prior) and delayed-tau fits converge to sub-solar metallicities, with all models suggesting that the metallicity of \ruby is 15-20\% solar. Our measurements of metallicity are indirect, in the sense that they are sensitive to the shape of the SED rather than specific spectral features, due to the low resolution of the PRISM spectrum and the relative weakness of metal-sensitive features as compared to the bright A-type stars that dominate the light of \ruby. However, these measurements are qualitatively consistent with the finding that massive quiescent galaxies at $z>1.4$ are metal poor relative to solar abundances \cite{Beverage2023,Beverage2024} and with comparably high redshift massive systems \cite{Carnall2023}. {On the other hand, because of a lack of very young metal-poor stars in the Milky Way, empirical model libraries suffer from calibration issues for young ages and low metallicities, which together with uncertainties in the stellar isochrones can have a significant effect on the shape of the continuum of the resulting SPS models \cite{Conroy2013,Whitler2023}.} We find that fits utilizing the C3K theoretical stellar libraries \cite{Conroy2012} rather than the empirical MILES libraries result in low-metallicity solutions within $\sim0.1$ dex of our fiducial runs and with comparable age measurements, highlighting that the preference of our fits to these low-metallicity solutions is insensitive to the choice of model libraries. { Importantly, however, these model libraries all assume solar abundance patterns. Recent results indicate that deviations in the elemental abundance patterns from the solar abundances used in the SPS models can also lead to incorrectly inferred metallicities\cite{Beverage2024}. Given that we measure very high \Nii/H$\alpha$, which can locally only be produced in AGN with high metallicity \cite{Groves2006}, we also fit the galaxy under the assumption of solar metallicity to quantify the effect that a model mismatch would have on the star formation history we infer.}

We find that fixing the metallicity to solar has affects the goodness of fit, as evidenced by the slightly worse $\frac{\chi^2}{N_\mathrm{data}}$ values of the fit (see Figure \ref{fig:model_comp}). However, this difference is largely driven by differences in the subtle shape of the rest-UV, where the NIRSpec PRISM resolution is worst and where stellar population libraries differ significantly. As such, it is difficult to reject this solution, even if models formally favor low metallicity solutions. The choice to fix the metallicity to solar has a negligible effect on the measured stellar mass, moving the median value by $\sim0.05$ dex, but has a considerable impact on the inferred star formation history. At solar metallicity, the observed spectrum and photometry of \ruby\ is best described by a stellar population that formed 50\% of its mass $\sim$400\,Myr later than the low-metallicity fiducial model. The fit still implies that galaxy assembled the majority of its mass within the first Gyr of cosmic time at the $2\sigma$ level, but the measured redshift where the galaxy had assembled 90\% of its mass (and can be considered to have quenched) shifts from $8.6\pm_{0.1}^{0.1}$ in the fiducial model to $5.7\pm_{0.0}^{0.1}$ in the fixed-metallicity model. The fixed-metallicity fit also implies a very similar burst-shape to the fiducial fit; despite the later-formation time, the estimated peak star formation rate, star formation duration, and star formation decline timescale (see the bottom panels in Figure \ref{fig:model_comp}) match quite closely between all models. Finally, the fixed-metallicity star formation history still implies an extremely efficient star formation history, requiring $\epsilon>0.2$ for the estimated number density of the source (see Figure \ref{fig:sfe}).  

\begin{table}
    \centering
    \caption{The resultant median and $1\sigma$ confidence intervals on a number of key galaxy parameters.}
    \begin{tabular}{cccccc}
    \toprule
     & Fiducial & Delayed-Tau & Rising & $Z=Z_\odot$ & n=1 \\
    \botrule
log$(M_\star/M_\odot)$ & $11.0\pm_{0.02}^{0.02}$ & $10.95\pm_{0.03}^{0.04}$ & $10.99\pm_{0.02}^{0.03}$ & $10.94\pm_{0.03}^{0.05}$ & $10.92\pm_{0.01}^{0.01}$ \\
\\
SFR$_{10}$ [$M_\odot$/yr] & $3.4\pm_{1.0}^{3.3}$ & $2.0\pm_{0.6}^{0.9}$ & $7.2\pm_{2.5}^{3.0}$ & $28.8\pm_{8.0}^{7.8}$ & $0.0\pm_{0.0}^{0.2}$ \\ \\
SFR$_{100}$ [$M_\odot$/yr] & $4.0\pm_{0.9}^{1.0}$ & $2.9\pm_{0.9}^{1.3}$ & $7.2\pm_{2.5}^{3.0}$ & $27.4\pm_{15.4}^{6.4}$ &$0.4\pm_{0.4}^{0.8}$ \\ \\
SFR$_{peak}$ [$M_\odot$/yr] & $870\pm_{140}^{70}$ & $470\pm_{40}^{60}$ & $910\pm_{100}^{140}$ & $660\pm_{160}^{270}$ & $260\pm_{30}^{60}$ \\ \\
$t_{form}$ [Myr] & $480\pm_{10}^{30}$ & $480\pm_{40}^{30}$ & $530\pm_{10}^{10}$ & $890\pm_{30}^{20}$ &$570\pm_{50}^{50}$ \\
\\
$z_{form}$ & $10.0\pm_{0.4}^{0.2}$ & $9.9\pm_{0.4}^{0.6}$ & $9.3\pm_{0.1}^{0.1}$ & $6.3\pm_{0.1}^{0.2}$ &$8.8\pm_{0.5}^{0.7}$ \\
\\
$t_{90}$ [Myr] & $590\pm_{0}^{10}$ & $710\pm_{30}^{20}$ & $590\pm_{0}^{0}$ & $1000\pm_{30}^{10}$ &$800\pm_{30}^{30}$ \\
\\
$z_{90}$ & $8.6\pm_{0.1}^{0.1}$ & $7.4\pm_{0.2}^{0.2}$ & $8.5\pm_{0.0}^{0.0}$ & $5.7\pm_{0.0}^{0.1}$ &$6.8\pm_{0.2}^{0.2}$ \\
\\
$t_{90} - t_{form}$ [Myr] & $100\pm_{10}^{10}$ & $230\pm_{20}^{20}$ & $60\pm_{10}^{10}$ & $110\pm_{20}^{20}$ &$230\pm_{30}^{40}$ \\
\\
$t_{90} - t_{10}$ [Myr] & $180\pm_{10}^{170}$ & $350\pm_{30}^{30}$ & $180\pm_{10}^{20}$ & $190\pm_{10}^{270}$ &$520\pm_{70}^{40}$ \\
\\
$A_\mathrm{V}$ [mag] & $0.17\pm_{0.05}^{0.05}$ & $0.13\pm_{0.06}^{0.11}$ & $0.18\pm_{0.05}^{0.08}$ & $0.48\pm_{0.13}^{0.06}$ &$0.06\pm_{0.01}^{0.02}$ \\
\\
log($Z/Z_\odot$) & $-0.73\pm_{0.02}^{0.02}$ & $-0.75\pm_{0.02}^{0.02}$ & $-0.74\pm_{0.02}^{0.02}$ & 0 &$-0.67\pm_{0.04}^{0.04}$ \\
        
        \botrule
    \end{tabular}
    \footnotetext[1]{SFR averaged over the 10 Myr over observation}{}
    \footnotetext[2]{SFR averaged over the 100 Myr over observation}{}
    \footnotetext[3]{The age of the universe when 50\% of the galaxy's mass formed}{}
    \footnotetext[4]{The age of the universe when 90\% of the galaxy's mass formed}{}
    \footnotetext[5]{$A_{\rm V}$ surrounding $t>10$ Myr stars. Our models assume $A_{\rm V}$ is doubled around $t<10$ Myr stars.}{}
    \label{tab:sfh_results}
\end{table}

Future deep observations that can directly measure metal-sensitive features and that are less beholden to the low resolution of the NIRSpec/PRISM will be able to place more robust constraints on the precise age, metallicity, and dust content of this system. However, despite uncertainty in the exact age of the stellar population, the observed PRISM spectrum and photometry unambiguously demonstrates that \ruby\ is a massive ($M_*\approx10^{11}\,\Msun$) quiescent galaxy that assembled the majority of its mass within the first $\sim$1 Gyr of cosmic time before rapidly quenching at $z>5.5${, which strains current theoretical models (see Section~\ref{sec:theory}).}

\section{Emission line fitting}\label{sec:eline_fitting}

\begin{figure}
    \centering
    \includegraphics[width=0.95\linewidth]{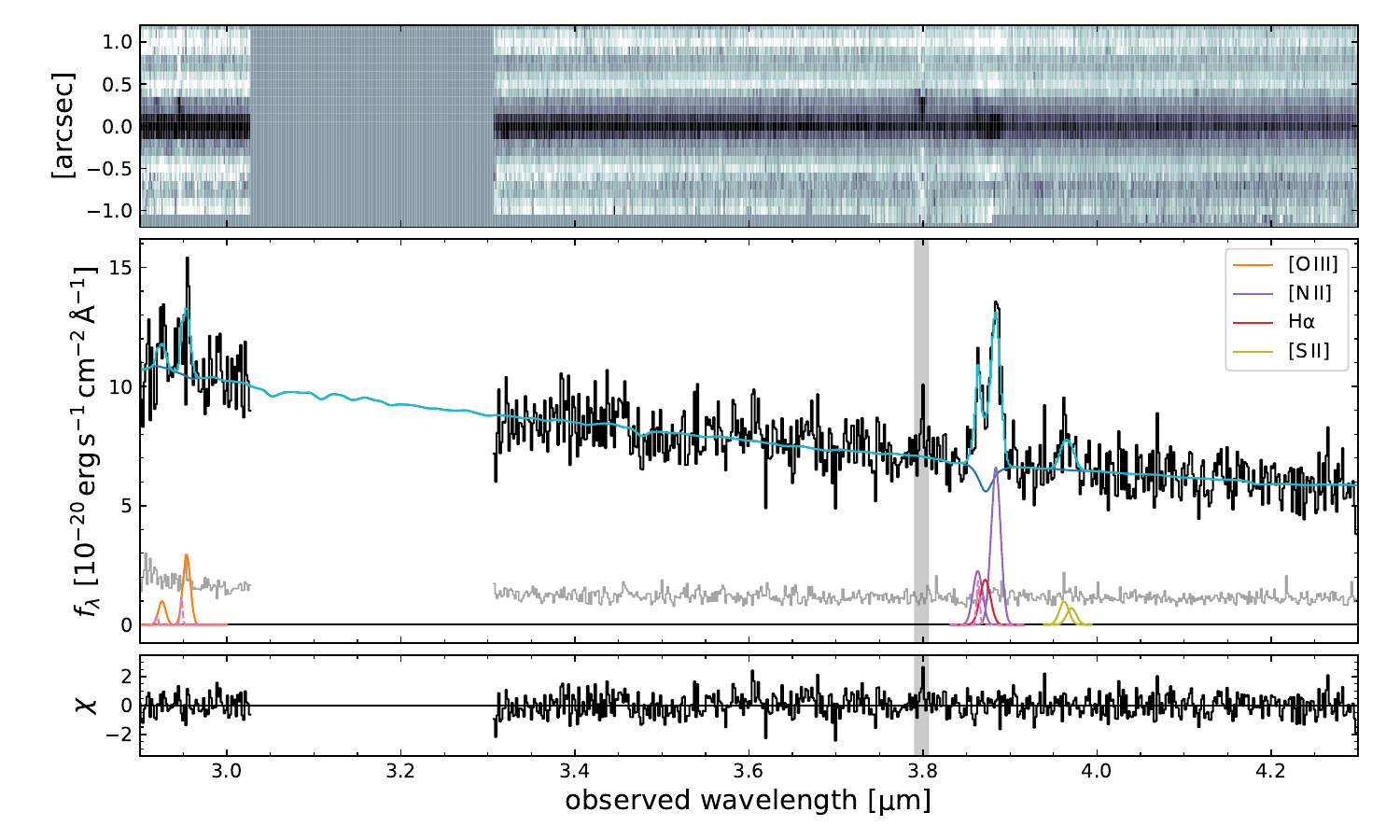}
    \caption{JWST/NIRSpec medium-resolution spectrum of \ruby, resolving the \Oiii, \Nii, and \Sii\ doublets, and the \Ha\ line. The $1\sigma$ uncertainties are show in gray. {The shaded region marks wavelengths that were masked in the fitting, as this emission originates from a source in a different slit in the MSA.} The dark blue line shows the median posterior model of the stellar continuum from \prospector. Colored lines show the median posterior models of the emission lines, and the cyan line shows the combined continuum and emission line model. {Emission lines originating from a satellite source in the same microshutter, apparent from their spatial offset in the 2D spectrum, are shown with pink dashed lines. }}
    \label{fig:g395m}
\end{figure}

The G395M spectrum obtained with JWST/NIRSpec (Figure~\ref{fig:g395m}) has a higher spectral resolution ($R \sim 1000-1500$) and resolves the emission line doublets and \Ha\ and \Nii\ complex that are blended in the PRISM spectrum. {We also find several emission lines that are spatially offset from the spectrum of \ruby. Crucially, we find that the lines at $2.95\,\micron$ and $3.85\,\micron$ lines are present in both the PRISM and G395M 2D spectra. We identify these lines as \Oiii\ and \Ha\ emission from the faint blue source found in the F115W image (Figure~\ref{fig:image_stamps}). Additionally, there is one emission line present in the spectrum at $3.80\,\micron$, which is not seen in the PRISM spectrum and implies that it originates from a different source on the NIRSpec MSA: the 2D spectra for the two sources overlap on the detector, because the G395M spectra are longer than the PRISM traces.} 

{We perform a simultaneous fitting to the continuum and \Oiii, \Nii, \Sii\ doublets. Although the extraction kernel used to obtain the 1D spectrum mitigates contamination from the satellite and spurious sources, some emission from these sources still appears. We therefore mask the contaminant emission line at $3.80\,\micron$, but explicitly include the \Oiii\ and \Ha\ lines of the satellite source in our model.} We model each emission line with a Gaussian line profile. The line ratio of \Nii\ is fixed to 1:2.94, {and that of \Oiii\ to 1:2.98}. Given the limited signal-to-noise ratio of the emission lines, we assume that all emission (and absorption) lines have the same velocity dispersion, which is a free parameter in the fit. We also include an \Ha\ emission line component, to estimate the infilling of the Balmer absorption line. As the \Ha\ emission may have a different origin from the other emission lines, we leave its velocity dispersion as an additional free parameter. To model the continuum, we use the deconvolved median posterior model from our \prospector\ fitting to the PRISM spectroscopy. We fit a {1st}-order polynomial between this continuum and the G395M spectrum to flux calibrate the spectrum. We assume all components {(except for the satellite source)} are at the same redshift.

The model is convolved with a custom line spread function (LSF), tailored to the morphology of \ruby\ based on the S\'ersic fit at $4\,\micron$\cite{deGraaff2023}. We allow for uncertainty in the LSF following the method described in \cite{Wang2024}. To estimate the posterior distributions of the parameters, we use the \texttt{emcee} package to perform Markov Chain Monte Carlo (MCMC) sampling. We adopt uniform priors for all parameters, allowing for velocity dispersions in the range $\rm \sigma_{\rm gas}\in[0,750]~\kms$. {We set a broader prior for the H$\alpha$ velocity dispersion, $\sigma_{\rm H\alpha}\in[0,1500]~\kms$, to test whether there is evidence for a broad line AGN.} %For the \Ha\ line we allow for negative flux to account for any mismatch between the \prospector\ model and data, and to avoid asymmetric uncertainties.

The emission line fluxes are reported in Table~\ref{tab:eline_fluxes}. We show the median posterior model of the combined continuum and emission line fitting in Figure~\ref{fig:g395m}, as well as the individual emission line and continuum components. {From the fitting we obtain a redshift of $z_{\rm spec}=4.8976_{-0.0010}^{+0.0006}$ and ionized gas velocity dispersion of $\sigma_{\rm gas}=414_{-64}^{+56}\,\kms$; we find that the velocity dispersion of the \Ha\ line converges to a similar value $\sigma_{\rm H\alpha}=461_{-150}^{+163}\,\kms$. The satellite source has a redshift of $z_{\rm spec}=4.8885_{-0.0009}^{+0.0009}$ and is thus offset by approximately $600\,\kms$ in the rest frame.}

We find very weak \Ha\ emission, marginally detected at the $2\sigma$ level. We hence measure the emission line ratio $\log($\Nii$_{\lambda6585}$/\Ha $)\,= 0.50_{-0.25}^{+0.34}$. Although the H$\beta$ line falls outside of the wavelength range covered by the G395M spectrum, we can obtain a lower limit on the H$\beta$ emission line flux by assuming case B recombination (i.e. \Ha/H$\beta =2.86$): in this case, the lower limit on the ratio $\log($\Oiii$_{\lambda5008}$/H$\beta) = 0.50_{-0.26}^{+0.31}$. These line ratios indicate that the line emission in \ruby\ does not originate from star formation\cite{Kewley2001}. Although shocked gas also results in high \Nii/\Ha\ line ratios, the ratio inferred from the spectrum and lower limit on the \Oiii/H$\beta$ ratio are most likely consistent with ionization by an AGN\cite{Veilleux1987,Kewley2006,Newman2018}.

%% Figure showing spectrum + fit
%% Table with all line fluxes
{We use the ionized gas velocity dispersion to estimate the dynamical mass using the methodology presented by \cite{vanDokkum2015} to relate the gas kinematics to the gravitational potential for compact quiescent galaxies. This assumes the gas forms a rotating disk, with a rotational velocity that is related to the integrated gas velocity dispersion and the inclination ($i$) of the disk: $v_{\rm rot} = \sigma_{\rm gas}/(\alpha \sin(i))$, where $\alpha\approx0.8$. Based on our morphological modeling (Section~\ref{sec:size}) we find that the projected axis ratio $q\approx0.85$, and assuming an intrinsic disk thickness of $q_0=0.2$ this implies $i=32.5$ degrees. The dynamical mass is computed as $M_{\rm dyn}= 2 v^2_{\rm rot}r_{\rm e} /G$, where $r_{\rm e}$ is the inferred half-light radius (Section~\ref{sec:size}), and $G$ the gravitational constant. With $r_{\rm e}=0.55\pm0.01$ kpc, we find $M_{\rm dyn} = 2.7_{-0.8}^{+0.7}\times 10^{11}\,\Msun$, a factor $\approx 3$ higher than the measured stellar mass. We note that the gas kinematics may be a biased tracer of the gravitational potential, as the observed stellar and ionized gas kinematics for a large sample of galaxies at $z\sim1$ have been shown to agree well on average\cite{Bezanson2018}, but with a systematic offset of approximately 0.2\,dex for $\sigma_{\rm gas}\sim 400\,\kms$. This may imply that the dynamical mass is overestimated by 0.4\,dex (a factor 2.5, $M_{\rm dyn} \approx 1.1\times10^{11}\,\Msun$), which is consistent with the estimated stellar mass.}

\begin{table}[h]
\caption{Emission line fluxes of \ruby\ measured from the G395M spectrum.}\label{tab:eline_fluxes}%
\begin{tabular}{@{}ll@{}}
\toprule
 & flux  \\
 & [$10^{-18}\,\rm erg\,s^{-1}\,cm^{-2}$] \\
\midrule
\Oiii$_{\lambda4960}$     &  $1.04_{-0.28}^{+0.31}$  \\
\Oiii$_{\lambda5008}$     & $3.09_{-0.83}^{+0.92}$  \\
\Nii$_{\lambda6549}$     & $3.09_{-0.51}^{+0.46}$  \\
\Ha     & $2.87_{-1.43}^{+1.45}$  \\
\Nii$_{\lambda6585}$     &  $9.09_{-1.49}^{+1.36}$ \\
\Sii$_{\lambda6718}$     &  $1.36_{-0.55}^{+0.91}$ \\
\Sii$_{\lambda6733}$     &  $0.88_{-0.54}^{+0.60}$ \\
\botrule
\end{tabular}
%\footnotetext{Source: This is an example of table footnote. This is an example of table footnote.}
\end{table}

\section{Size measurement}\label{sec:size}

We measure the effective radius ($r_{\rm e}$) of \ruby\ as a function of wavelength using the GALFIT \citep{Peng02,Peng10} single component S\'ersic fitting methods from \cite{vanderWel12} in all available filters (F115W, F150W, F200W, F277W, F356W, F410M, and F444W), corresponding to rest-frame wavelengths in the range $\sim2000$ to 7500 $\mathrm{\AA}$. 

All filters are processed following the procedure described in \cite{Cutler24}. 
% A cutout is created of \ruby\ from the science, weight, exposure time, and segmentation images with a length of $7\times R_{\rm{Kron,circ}}$ (234 pixels at 0.04 $^{\prime\prime}$/pixel scale) for each filter. The error at a given pixel, $\sigma_i$, is estimated using a combination of sky background variance ($1/w$) and Poisson noise ($f/t_{\rm exp})$, following Equation 4 in \cite{Cutler24}. Nearby sources within the cutout are masked using the segmentation map. We use empirical JWST PSFs that are built using stacks of unsaturated stars selected directly from the same data set, as in \cite{Weaver24}, normalizing to reported encircled energies from calibration resources\footnote{\url{https://jwst-docs.stsci.edu/jwst-near-infrared-camera/nircam-performance/nircam-point-spread-functions}} at 4$^{\prime\prime}$.  
% The science, error, and mask cutouts, as well as the empirical PSFs, are provided to GALFIT and used to determine the best-fit S\'ersic model.  The free parameters include the object centroid ($x_0$, $y_0$), total magnitude ($m$), semi-major axis effective (half-light) radius ($r_e$), S\'ersic index ($n$), axis ratio ($q$), and position angle ($\theta$). Following \cite{Cutler24}, we do not fit for an additional sky background pedestal, as JWST performs better with external background subtractions. 
Parameters are constrained in the model fits as follows: the magnitude can range $\pm3$ mag from the photometric catalog value, radius varies from $0.01<r_{\rm e}<400$ pixels, S\'ersic index can range from $n=0.2$ to $n=10$, and axis ratio from $q=0.0001$ (flat) to $q=1$ (round).

Next, sizes are corrected to account for residual flux using the methods of \cite{Szomoru2010}. The growth curve from the best-fit S\'ersic model (deconvolved from the PSF) is added to the GALFIT residual growth curve. We extrapolate the combined growth curve using the S\'ersic model alone when the annular S/N$<3$ for the science image. The corrected radius is then defined as the radius where the residual$+$unconvolved S\'ersic growth curve reaches 50\%.

\begin{figure}
    \centering
    \includegraphics[width=\linewidth]{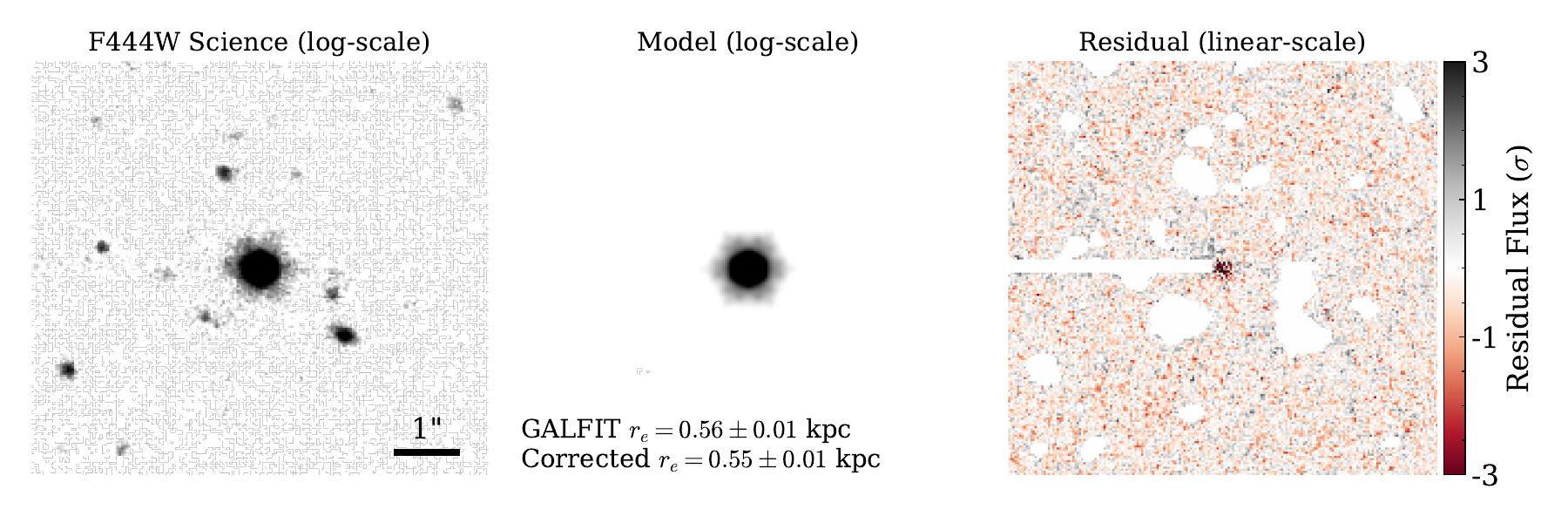}
    \caption{S\'ersic profile fitting for the F444W image reveals a compact light distribution ($n\approx9$, $r_{\rm e}\approx0.55\,$kpc).}
    \label{fig:morph}
\end{figure}

We find that \ruby\ is remarkably compact, although we unambiguously resolve this galaxy in F356W, F410M, and F444W. The F444W fit is shown in Figure~\ref{fig:morph}, with a very compact profile ($n=8.80\pm0.14$) and a corrected half-light radius of $r_{\rm e}=0.55\pm0.01\,$kpc (consistent with \cite{Ito2023}). The F356W and F410M fits find similar S\'ersic index (8.77 and 8.15, respectively) and corrected radii (0.50 and 0.58 kpc), indicating a flat color gradient in the rest-frame optical (3400$\mathrm{\AA}$ to 7500$\mathrm{\AA}$). Given the rapid assembly and quenching of \ruby\, it is not surprising to find a flat color gradient \citep[see, e.g.,][]{Suess2020}. The shorter wavelength images (rest-frame $<3000\mathrm{\AA}$) are very compact and have trouble converging to accurate fits without hitting the $n=10$ upper limit.

% if we fix $n=4$ for F115W, F150W, F200W, and F277W. For the F356W, F410M, and F444W images we are able to successfully model the light profiles with the S\'ersic index as a free parameter, finding very compact profiles with $n\approx9$ and a corrected half-light radius of $r_{\rm e}=0.56\pm0.02\,$kpc in the F444W band. We also find that it has a flat color gradient in the rest-frame optical, from 3400$\mathrm{\AA}$ to 7500$\mathrm{\AA}$ (Figure~\ref{fig:morph}; though it is much smaller in the rest-ultraviolet, $2000\mathrm{\AA}$ to $3000\mathrm{\AA}$).

\section{Environment at $z=4.9$}

% relevant papers include Zavala+, Arabal-Haro+, Naidu+, ...$
% spec-z's come from RUBIES (193), CEERS (123), DDT (20) with grade=3 from DJA -- of these, 289 have NIRCam counterpart
% some mention of velocity uncertainty in prism vs grating, allow for systematic uncertainty to search for velocity clustering
% describe estimate of background clustering and how this is a 3-4 sigma detection wrt this background
Recent studies have speculated on the existence of an overdensity at $z\approx5$ in the EGS field, based on a clustering of photometric redshifts\cite{Naidu2022}, and a sample of 4 spectroscopic redshifts at $z\approx4.90$ obtained with JWST/NIRSpec\cite{Arrabal2023}. With the addition of \ruby, this gives a sample of 5 spectroscopic redshifts.

To search for further evidence of a large-scale overdensity, we leverage all available spectroscopic redshifts from JWST/NIRSpec in the EGS field. These spectra come from different JWST Cycle 1 and Cycle 2 programs: the CEERS program, a Director's Discretionary Time program (DDT, \#2750; PI Arrabal Haro), and the RUBIES program. All data on the DJA were reduced using the \texttt{msaexp} pipeline\cite{Brammer2022}, in the same manner as described in Section~\ref{sec:nirspec}. Redshifts were obtained from the spectra using $\chi^2$ minimization template fitting with \texttt{msaexp}, and visually inspected to evaluate the quality of the redshifts. The redshifts are predominantly derived from PRISM spectroscopy, although approximately half of the sources in CEERS were observed with the medium-resolution gratings only. We begin by compiling all sources with robust redshifts (grade\,$=3$ from visual inspection) in the range $4<z<6$ from the DJA, which results in 193 sources from RUBIES, 123 from CEERS and an additional 20 from the DDT program. Of these 336 sources, 289 are detected in NIRCam imaging, with the remainder falling outside of the NIRCam footprint.

\begin{figure}
    \centering
    \includegraphics[width=\linewidth]{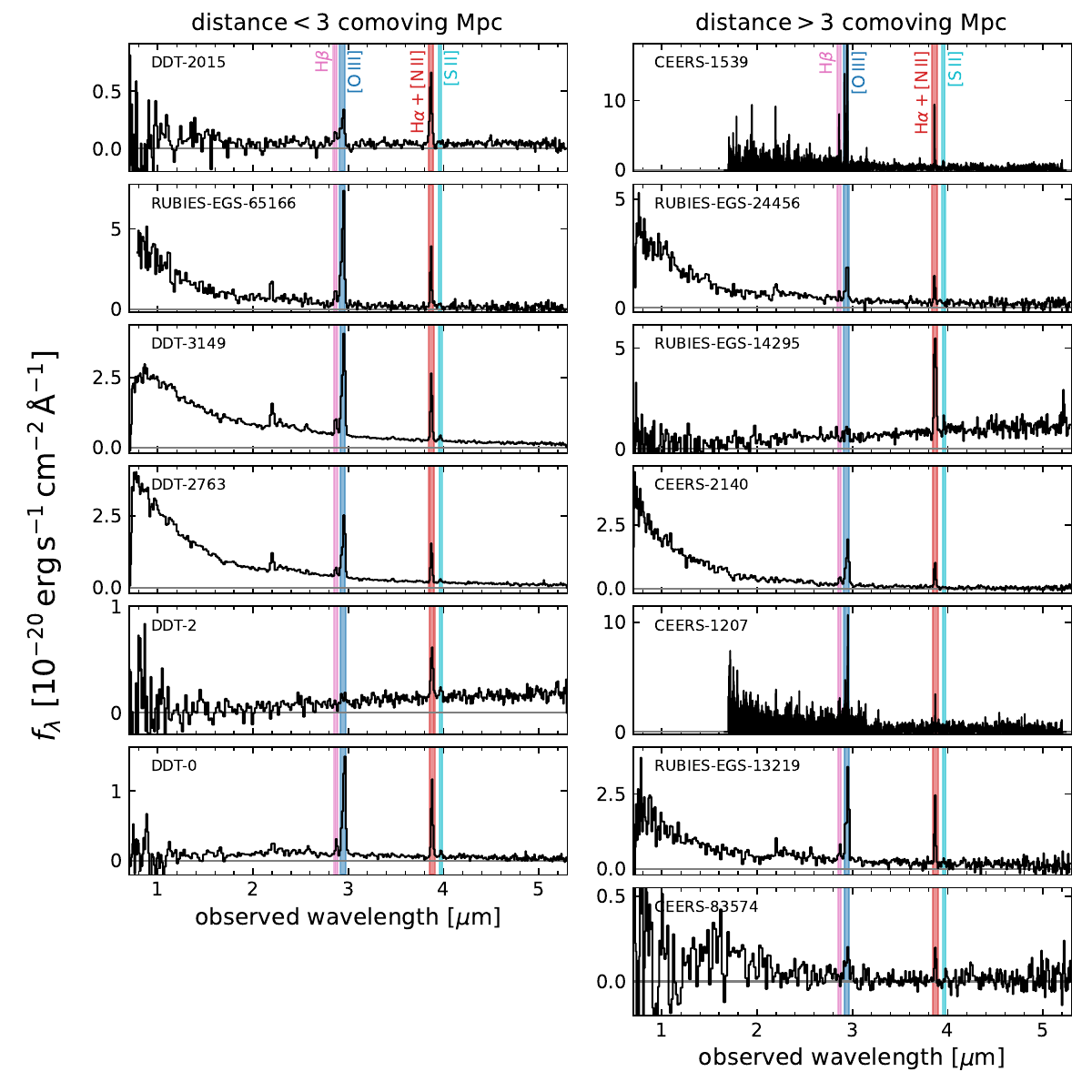}
    \caption{JWST/NIRSpec spectra of 13 sources with a redshift $z\approx4.90$. The left column shows objects that have a close projected separation to \ruby, in order of increasing redshift. The right column shows sources in the EGS that are at a larger distance from \ruby. }
    \label{fig:z490_spectra}
\end{figure}

We select galaxies within {$\Delta z= \pm 0.0135$ ($\Delta v = 4000\,\kms$)} from the redshift of \ruby, chosen based on the typical redshift ranges used to search for overdensities (from the compilation in \cite{Harikane2019}), which is sufficiently large to account for systematic uncertainties in the wavelength calibration of NIRSpec\cite{Bunker2023}. In total, we find 13 sources in this velocity range around \ruby\ (Figure~\ref{fig:z490_spectra}, Table~\ref{tab:overdensity}), 4 of which were published previously. Of these 13 sources, 6 have a close angular separation to \ruby\ of $<1$\,arcmin, corresponding to $<400\,$kpc or $<2\,$ comoving Mpc. We identify another clustering of 4 sources at a distance of $\approx7$\,arcmin ($\approx 16$ comoving Mpc) from \ruby. Interestingly, the brightest source in the F444W imaging among these four (RUBIES-EGS-14295) is a submillimeter galaxy identified from SCUBA-2 $850\,\micron$ data (S2CLS-EGS-850.061; \cite{Zavala2017}), and implies a high star formation rate. The remaining three sources are scattered across the field at varying distances.

The typical NIRSpec MSA can target approximately $\sim 200$ sources simultaneously, which generally leads to a complex selection function, in which high-priority targets have a high probability of being allocated a shutter, and any individual galaxy in the broader population has a low probability of being observed\cite{Ferruit2022,Bunker2023,Maseda2024}. As a result, the spectroscopic completeness is likely to be low at intermediate redshift, as the science objective of the majority of programs focuses on galaxies at $z>6$. Because of this incompleteness, it is difficult to quantify the overdensity of the spectroscopic targets at the redshift of \ruby\ to estimate its halo mass. However, we can use the $4<z<6$ galaxy population with robust spectroscopy to assess qualitatively whether \ruby\ is likely to be part of an overdensity. Under the assumption that sources in this redshift range are targeted with approximately equal probability, we can estimate the the typical clustering of galaxies in redshift space and projected distance. 

To assess the expected clustering in {redshift} space, we select the 50 brightest sources in F444W NIRCam imaging among the 336 spectroscopic targets. We then search for sources within a radius of 3 comoving Mpc, approximately 3 times the virial radius of a moderately-sized cluster at $z=0$ ($\rm M_{halo}\sim 10^{14.5}\,\Msun$), and compute the {redshift} separation between each source and the brightest galaxy. The corresponding distribution in {redshift} space peaks at $\Delta z=0$, but is very broad (Figure~\ref{fig:environment}). Within a range of {$\Delta z=0.027$}, we find that the overdensity around \ruby\ is significantly ($3.1\sigma$) above this background level. If we use a larger aperture, we find only marginal detections of $2.2\sigma$ (5\,Mpc), $1.5\sigma$ (10\,Mpc) and $1.6\sigma$ (20\,Mpc).

Similarly, we select all sources in narrow redshift windows of $\Delta z=0.027$, stepping between $z=4.0$ to $z=6.0$, and measure the angular separation with respect to the brightest source in the redshift slice. We again find a broad distribution in the projected distance, which we use to estimate the significance of the clustering of sources around \ruby. Within a radius of $3\,\Mpc$, the neighbors of \ruby\ are $2.8\sigma$ above the expected level. As before, this decreases when searching within larger apertures: $2.0\sigma$ within $5\,\Mpc$; $1.0\sigma$ within $10\,\Mpc$; $1.44\sigma$ within $20\,\Mpc$.

We conclude that \ruby\ indeed resides in an overdense environment, at least within an aperture of 3 comoving Mpc. Currently this forms the highest redshift overdensity containing a massive quiescent galaxy\cite{Kakimoto2024}. %By extrapolating the stellar-halo mass relation\cite{Moster2018,Behroozi2019}, we can obtain an approximate estimate for the halo mass of $\sim10^{12.5}\,\Msun$, which would typically grow to a galaxy cluster of halo mass $\sim 10^{14-14.5}$ at $z=0$. 
%Possibly, this implies accelerated growth of massive quiescent galaxies in overdensities, although such overdensities at high redshift do not necessarily form clusters at low redshift\cite{Remus2023a}. 
However, based on these findings we cannot determine whether \ruby\ resides in a massive halo or in the extremely massive halo needed to reconcile the observed number density of the galaxy with expectations from galaxy formation simulations. Further spectroscopic follow-up observations will be critical to perform a thorough analysis of the spectroscopic completeness in order to estimate a robust halo mass at the observed redshift, and the projected evolution to the present day.

\begin{table}[h]
\caption{Spectroscopically-confirmed sources in the EGS field at the redshift of \ruby.}\label{tab:overdensity}%
\begin{tabular}{@{}llll@{}}
\toprule
ID & R.A. & Dec. & $z$  \\
\midrule
CEERS-1207\footnotemark[1] & 214.960005 & 52.831171 & 4.8957 \\
CEERS-1539\footnotemark[1] & 214.980078 & 52.942659 & 4.8840 \\
CEERS-2140\footnotemark[1] & 214.796009 & 52.715878 & 4.8927 \\
CEERS-83574 & 214.949862 & 52.831306 & 4.8980 \\
DDT-0\footnotemark[2] & 214.914550 & 52.943023 & 4.9098 \\
DDT-2\footnotemark[2]$^,$\footnotemark[3] & 214.909113 & 52.937204 & 4.9080 \\
DDT-2015 & 214.917995 & 52.937245 & 4.8903 \\
DDT-2763\footnotemark[2] & 214.927789 & 52.935859 & 4.9009 \\
DDT-3149\footnotemark[2] & 214.914917 & 52.943621 & 4.8991 \\
RUBIES-EGS-13219 & 214.947589 & 52.836578 & 4.8966 \\
RUBIES-EGS-14295 & 214.943835 & 52.835816 & 4.8925 \\
RUBIES-EGS-65166 & 214.918350 & 52.931829 & 4.8919 \\
\botrule
\end{tabular}
\footnotetext[1]{Published in \cite{Nakajima2023}.}
\footnotetext[2]{Published in \cite{Arrabal2023}.}
\footnotetext[3]{CEERS-DSFG-1 of \cite{Zavala2023}.}
\end{table}

\section{Theoretical predictions}\label{sec:theory}

We compare the existence and properties of \ruby\ with {five} different large-volume simulations from {recent} literature, all of which simulate galaxy formation and evolution within the $\Lambda$CDM cosmological model. First, we evaluate the number density of massive quiescent galaxies in the different models: 
{
\begin{itemize}
    \item The First Light And Reionisation Epoch Simulations (FLARES)\cite{Vijayan2021,Lovell2021} uses the EAGLE model\cite{Crain2015,Schaye2015} to perform hydrodynamical zoom simulations of regions in a volume of 3.2 comoving Gpc$^3$, thereby probing rare haloes that would not appear in the EAGLE simulation. Using the measurements of \cite{Lovell2023} based on the combination of FLARES and EAGLE, the predicted surface density of quiescent galaxies at $z\sim4.5-5$ and apparent magnitude similar to \ruby\ ($\rm F200W<24.5$) is approximately $0.1-1\,$degree$^{-2}$, or $n\sim 1-10\times10^{-8}\,\Mpc^{-3}$. 
    \item In the cosmological hydrodynamical simulation TNG300 of the IllustrisTNG project\cite{Nelson2018a,Nelson2019a,Pillepich2018b,Springel2017,Marinacci2018,Naiman2018}, which simulates a comoving volume of $302\,\Mpc^3$ using the IllustrisTNG model, the population of massive quiescent galaxies has been shown to appear only at $z\approx4.2$\cite{Hartley2023} in the simulation. At $z=5$, we find a single system in the simulation that lies within the $3\sigma$ contours of the stellar mass and SFR of \ruby\ (as measured within an aperture of twice the 3D stellar half-mass radius), corresponding to a comoving number density of $n=3\times10^{-8}\,\Mpc^{-3}$.
    \item  In the cosmological hydrodynamical simulation Magneticum Pathfinder of comoving volume $180\,\Mpc^3$, lower-mass quiescent galaxies are present with number densities consistent with observations\cite{Kimmig2023}, but no systems were reported at $M_*\sim10^{11}\,\Msun$ at $z\approx5$, setting an upper limit on the number density of $n< 1\times10^{-7}\,\Mpc^{-3}$.
    \item  The latest version of the semi-analytic GAlaxy Evolution and Assembly (GAEA) model\cite{Hirschmann2016,Fontanot2020,Xie2020,DeLucia2024} was run on the Millennium Simulation\cite{Springel2005} of box length $500\,\Mpc/h$ (where $h=0.73$). Within this large volume there are 14 quiescent galaxies of $\log(M_*/\Msun)>10.9$ at $z\approx 4.9$, i.e. $n=4\times10^{-8}\,\Mpc^{3}$.
    \item The semi-analytic model SHARKv2.0\cite{Lagos2024} was run for the SURFS\cite{Elahi2018} dark matter-only simulation volume of box length $210\,\Mpc/h$ (with $h=0.6751$). Following the methodology of \cite{Lagos2024}, which includes an estimate of the observational uncertainty of $0.25\,$dex on the inferred number density, the number density of massive ($\log(M_*/\Msun)>10.9$) quiescent galaxies $n=2\times10^{-8}\,\Mpc^{3}$ at $z=5$.
\end{itemize}
}

In summary, we find that the different models all predict the existence of massive quiescent galaxies at high redshifts, but with very low number densities at the redshift of \ruby, {with a typical value of $n\sim 5\times 10^{-8}\,\Mpc^{-3}$ across the simulations}. %The small area targeted with JWST spectroscopy in the RUBIES survey ($\approx100$\,arcmin$^2$ thus far) implies an observed number density of $n\approx3\times10^{-6}\,\Mpc^{-3}$ (at $4.5<z<5.5$), and based on the photometric selection of quiescent galaxies more massive than $M_*>10^{10.9}\,\Msun$ at $4<z<5$ we obtain a number density of $n\approx 5\times 10^{-6}\,\Mpc^{-3}$\cite{Valentino2023}. 
{With an estimated observed number density of $n\approx4\times10^{-6}\,\Mpc^{-3}$,} \ruby\ is therefore an outlier at the $2\sigma$ level {at $z\approx5$} if we assume the theoretical model predictions to be accurate. It also implies that, instead of residing in a halo of $\Mhalo\sim10^{12}\,\Msun$, the halo mass of \ruby\ would be among the most massive haloes at $z=5$, as $\Mhalo\sim10^{12.5-13}\,\Msun$ in the simulations.

{However, at higher redshift the tension increases: even for the solar-metallicity model, \ruby\ formed its stars and quenched at $z\gtrsim5.5$. The FLARES simulation predicts an extremely low source density of $n < 1\times10^{-8}\,\Mpc^{-3}$ at $z\geq5.5$ (i.e. $\lesssim0.1\,$degree$^{-2}$ ), and the GAEA simulations predict similarly low numbers ($n=6\times10^{-9}\,\Mpc^{-3}$ at $z\approx6$); both report zero such galaxies at $z\sim7$, which implies that $n(z=7)\lesssim1\times10^{-9}\,\Mpc^{-3}$. The relatively smaller volumes (TNG300, Magneticum Pathfinder) do not appear to contain any quiescent galaxies more massive than $M_*>10^{10.9}\,\Msun$ at $z\geq5$, and do not contain any such massive galaxies at $z>6$ even when considering star-forming galaxies. %Comparing the observed and simulated number densities implies that \ruby\ is an outlier at the $>2.7\sigma$ level. %Assuming standard recipes for star formation implemented in the different models, this indicates that \ruby\ must have formed in an exceptionally massive halo at $z\gtrsim7$.
Therefore, if we also account for the unusual formation history of \ruby, this implies that the probability of finding a source like \ruby\ within the small volume of the NIRCam imaging explored for the RUBIES survey thus far is approximately 0.03\% (i.e., an outlier at the $>2.8\sigma$ level), and strains current theoretical models.
}

Lastly, we use the GAEA and TNG300 simulations to explore the possibility that \ruby\ formed through a rapid, major merger of two moderately-massive galaxies. We estimate a merger timescale of $\sim 200-300\,$Myr\cite{Solanes2018}: because we do not find any signatures of a recent merger in the NIRCam imaging, the merger must have occurred at $z>6$. We therefore search for the nearest massive neighbor of galaxies of $\log(M_*/\Msun)>10.3$ in the simulations at $z\gtrsim 6$. For GAEA, we find two massive pairs that are separated by $<0.5\,$comoving Mpc at $z=6.2$, and six such pairs at $z=5.7$, corresponding to a number density of $6-20\times10^{-9}\,\Mpc^{-3}$. In TNG300, we find a single pair of massive galaxies at $z=6.0$, which implies a number density for such equal-mass mergers of $3\times10^{-8}\,\Mpc^{-3}$. However, when tracing their merger history, we find that the pair does not merge until $z=5.0$, which is inconsistent with the lack of merger signatures in the morphology of \ruby. These low number densities are therefore upper limits on the expected rate of equal-mass mergers of massive galaxies at high redshift, and indicates that the formation of \ruby\ through such a scenario is expected to be extremely rare.

\section{NOEMA non-detection}

\ruby\ was also covered by NOEMA observations as part of project W20CK (PIs: Buat \& Zavala), originally designed to target dusty star-forming galaxy candidates at $z>3$  (see \cite{Zavala2023} for further details and the data reduction process). These observations allow us to derive a 3$\sigma$ upper limit at 1.1mm of $\rm <1\,mJy$. Assuming a typical SED -- a modified black-body function with a dust temperature of $T_{\rm D}=35\,$K and a dust emissivity index of $\beta=1.8$ -- at $z=4.9$, this flux density upper limit corresponds to an IR luminosity of $<8\times10^{11}\,L_\odot$\,. Assuming the calibration between star formation rate and infrared luminosity of \cite{Kennicutt2012}, this implies a dust-obscured star formation rate of $<120\,\Msun\rm\,yr^{-1}$.

\bmhead{Acknowledgments}

%Acknowledgements are not compulsory. Where included they should be brief. Grant or contribution numbers may be acknowledged.

We thank V. Buat, D. Burgarella, and J. Zavala for sharing their NOEMA data and constraints on the dust-obscured star formation of RUBIES-EGS-QG-1. This work is partially based on observations carried out under project number W20CK with the IRAM NOEMA Interferometer. IRAM is supported by INSU/CNRS (France), MPG (Germany), and IGN (Spain). {We thank Claudia Lagos for providing measurements from the SHARK simulation.} 
This research was supported by the International Space Science Institute (ISSI) in Bern, through ISSI International Team project \#562.
MVM, JL, and BW acknowledge funding support from NASA via JWST-GO-4233. The Cosmic Dawn Center is funded by the Danish National Research Foundation (DNRF140). This work has received funding from the Swiss State Secretariat for Education, Research and Innovation (SERI) under contract number MB22.00072, as well as from the Swiss National Science Foundation (SNSF) through project grant 200020\_207349. Support for this work was provided by The Brinson Foundation through a Brinson Prize Fellowship grant. {Support for this work for RPN was provided by NASA through the NASA Hubble Fellowship grant HST-HF2-51515.001-A awarded by the Space Telescope Science Institute, which is operated by the Association of Universities for Research in Astronomy, Incorporated, under NASA contract NAS5-26555.} {This work is based on observations made with the NASA/ESA/CSA James Webb Space Telescope. The data were obtained from the Mikulski Archive for Space Telescopes at the Space Telescope Science Institute, which is operated by the Association of Universities for Research in Astronomy, Inc., under NASA contract NAS 5-03127 for JWST. The observations in this work are associated with programs ERS-1345, GO-2234, DDT-2750 and GO-4233. We gratefully acknowledge the CEERS and DDT-2750 teams for developing their observing program with a zero-exclusive-access period.}

\bmhead{Data availability}

The unprocessed JWST data are available through the Mikulski Archive for Space Telescopes. Reduced data from JWST underlying this work are publicly available in the DAWN JWST Archive (\url{https://dawn-cph.github.io/dja/}). 

\bmhead{Code availability}

All software packages used in this work  are publicly available on Github: \texttt{grizli}, \texttt{msafit}, \texttt{msaexp}, \prospector, \texttt{sedpy}. The \textsc{Galfit} software can be found on \url{https://users.obs.carnegiescience.edu/peng/work/galfit/galfit.html}.

\bmhead{Author contributions}

AdG and GB designed the NIRSpec MSA observations. AdG drafted the main text and led the emission line, environment, and simulation analyses. DJS led the stellar population analysis. GB performed the image and spectroscopic data reduction. SC led the morphological analysis. AW and IL measured the JWST and HST photometry. KAS, JL and MVM contributed to the stellar population analysis. KAS, SP and KEW contributed to the morphological analysis. FV provided the number density estimate. GDL and MH provided measurements from the GAEA simulation. All authors aided in the interpretation of the results and contributed to the Main text and Methods.

\bmhead{Author Information} The authors declare that they have no competing financial interests. Correspondence and requests for materials should be addressed to AdG (email: degraaff@mpia.de). 

\bibliography{bibfile}% common bib file

\end{document}